\documentclass[11pt,journal,twocolumn]{IEEETran}
\ifCLASSOPTIONcompsoc
  \usepackage[nocompress]{cite}
  \else
  \usepackage{cite}
  \fi
  \ifCLASSINFOpdf
  \else
\fi
\usepackage{multirow} 
\usepackage{algpseudocode}
\usepackage{algorithm}
\usepackage{rotating}
\usepackage{kantlipsum} 
\usepackage{commath}
\allowdisplaybreaks
\usepackage{mathtools}  
\usepackage{verbatim}
\usepackage{xr-hyper} 
\usepackage{epstopdf}
\usepackage{wrapfig}
\usepackage{latexsym}
\usepackage{amssymb}
\usepackage{amsthm}
\usepackage{amsfonts}
\usepackage{amsmath} 
\usepackage{graphicx}
\usepackage{latexsym}
\usepackage{booktabs}
\usepackage{subcaption} 
\algnewcommand\algorithmicinput{\textbf{INPUT:}}
\algnewcommand\INPUT{\item[\algorithmicinput]}
\algnewcommand\algorithmicoutput{\textbf{OUTPUT:}}
\algnewcommand\OUTPUT{\item[\algorithmicoutput]}
\usepackage[table]{xcolor}

\begin{document}
\title{SDN-Based Resource Allocation in MPLS Networks: A Hybrid Approach}


		
\author{Mohammad Mahdi Tajiki, Behzad Akbari, Nader Mokari, Luca Chiaraviglio
\IEEEcompsocitemizethanks{\protect
\IEEEcompsocthanksitem Luca Chiarviglio is with the ECE Department, University of Rome Tor Vergata, Italy, E-mail: luca.chiaraviglio@uniroma2.it\protect
\IEEEcompsocthanksitem MM. Tajiki, B. Akbari, N. Mokari are with the ECE Department, University of Tarbiat Modares, Tehran, Iran - E-mail: \{mahdi.tajiki,b.akbari,nader.mokari\}@modares.ac.ir
}}

\maketitle

\begin{abstract}
The highly dynamic nature of the current network traffics, makes the network managers to exploit the flexibility of the state-of-the-art paradigm called SDN. In this way, there has been an increasing interest in hybrid networks of SDN-MPLS. In this paper, a new traffic engineering architecture for SDN-MPLS network is proposed. To this end, OpenFlow-enabled switches are applied over the edge of the network to improve flow-level management flexibility while MPLS routers are considered as the core of the network to make the scheme applicable for existing MPLS networks. The proposed scheme re-assigns flows to the Label-Switched Paths (LSPs) to highly utilize the network resources. In the cases that the flow-level re-routing is insufficient, the proposed scheme recomputes and re-creates the undergoing LSPs. To this end, we mathematically formulate two optimization problems: i) flow re-routing, and, ii) LSP re-creation and propose a heuristic algorithm to improve the performance of the scheme. Our experimental results show the efficiency of the proposed hybrid SDN-MPLS architecture in traffic engineering superiors traditionally deployed MPLS networks.
\end{abstract}

\IEEEpeerreviewmaketitle

\section{Introduction}
\IEEEPARstart{S}{ervice} providers around the world have large investments in highly sophisticated and feature-rich MPLS network infrastructures for providing services to their customers. These infrastructures are built on traditional network equipment (combined data plane and control plane) which are costly to scale, complex to manage, and time consuming to reconfigure. Network Function Virtualization (NFV), cloud computing and the proliferation of connected devices are leading to exponentially increasing traffic and significant fluctuations in usage patterns. These reasons make network operators to move to agile architectures which support dynamic reconfiguration of both services and the network infrastructures \cite{tu2014splicing}. For Service Providers, these capabilities provide new revenues, reduce time to market, increase new service uptake, and enhance their ability to meaningfully differentiate their offerings\cite{KirkpatricSDN}.

The state-of-the-art paradigm called SDN \cite{SDNFirstPaper} along with the OpenFlow protocol \cite{OpenFlowProtocol} provides lots of new traffic management features\cite{FarhadiSDNSurv}. This makes it a proper and highly adopted technology for data center networks. One of the most important benefits of employing OpenFlow is its ability to route/re-route the traffic flows based on the network traffic pattern. In other words, it optimally routes/re-routes the traffic in flow level granularity. Therefore, there are lots of novel works which focus on traffic engineering in pure OpenFlow networks\cite{GholCong,TajikQRTP,AkyildizSDNRoadMap,TajikSDTE,tajiki2017MDPI}. However, migration from carrier networks which are mostly MPLS-based to OF-based network is challenging and highly expensive. 

To circumvent the aforementioned challenges, we propose a novel traffic engineering architecture in which the integration of OpenFlow and traditional MPLS is adopted. This traffic engineering architecture is motivated by scenarios where SDN is going to be deployed in an existing network. In such a network, some parts of the traffic is controlled by the SDN controller; some other parts of the network use existing network routing protocol. In other words, we consider traffic engineering in the case where a SDN controller controls only a few SDN forwarding elements in the network and the rest of the network does hop-by-hop routing using MPLS protocol. The objective is to propose a traffic engineering algorithm for integration of MPLS and OpenFlow networks that can adaptively and dynamically manage traffic in a network to accommodate different traffic patterns. To this end, the network traffic is monitored  to achieve the current traffic matrix. Thereafter, based on the current traffic matrix and the knowledge base of previous demands, the controller computes LSPs and assign the flows to each LSP (at the edge layer: OpenFlow-enabled switches). Our main contributions are as follows:
\begin{itemize}
\item A new traffic engineering architecture for MPLS-OpenFlow hybrid networks is proposed.
\item  We mathematically formulate two optimization problems: a) the problem of LSPs re-configuration in MPLS networks when there is a central controller as the PCE element, and b) the problem of flow-level resource re-allocation.
\item In order to improve the performance of the solution, a heuristic algorithm for the problem of flow-level resource re-allocation is proposed.
\end{itemize}

The remainder of the paper is organized as follows: In Section~\ref{relatedWork}, the related work is discussed. Section~\ref{ProposedArchitecture} states the definition of the problem, the proposed architecture, and an outline of the proposed schemes. Section~\ref{problemForumlation} discusses the system model, parameters, objective function and constraints. The proposed heuristic algorithm is described in Section~\ref{sec:hueristic}. The performance analysis of the proposed schemes are presented in Section~\ref{resultAnalysis}. Finally, Section~\ref{conclusion} concludes the paper and presents future directions.

\section{Related Works}\label{relatedWork}
In the following, we explain the state-of-art algorithms which are related to hybrid networks. To this end, we categorize the subject into three sub-topics: i) hybrid approaches that allows the coexistence of traditional IP  routing  and  SDN  based  forwarding  within  the  same  provider  domain, ii) hybrid approaches that focus on combination  of  traffic  engineering  and  power management in hybrid networks, and iii) incremental  deployment  of  hybrid  networks.

\subsection{IP routing and SDN based forwarding within the same provider domain}
Salsano et al. \cite{salsanoHybridIPSDN}, propose a hybrid approach that allows the coexistence of traditional IP routing with SDN based forwarding within the same provider domain. To this end, they design a hybrid IP/SDN architecture called Open Source Hybrid IP/SDN (OSHI). Besides, they implement a hybrid IP/SDN node made of Open Source components.
The aim of \cite{lopexTwardTransSDN} is to present some architecture to enable interoperability in transport networks. They present alternatives to control plane interoperability. Moreover, they justify why SDN can be a solution to enable multi-vendor scenario and multi-domain path establishment in current networks.
In \cite{AguadoABNO}, an application-based network operations (ABNO) architecture is proposed as a framework that enables network automation and programmability. ABNO not only justifies the architecture but also presents an experimental demonstration for a multi-layer and multi-technology scenario.

Sgambelluri et al. \cite{SgambelSDNPCE}, present two segment routing (SR) implementations for MPLS and SDN-based networks, separately. They have two different network testbeds. The first implementation focuses on a SDN scenario where nodes consist of OpenFlow switches and the SR Controller is an enhanced version of an OpenFlow Controller. The second implementation includes a Path Computation Element (PCE) scenario where nodes consist of MPLS routers and the SR Controller is a new extended version of a PCE solution. 

Das et al. \cite{DasMPLSOPN}, propose an approach to MPLS that uses the standard MPLS data plane and  an OpenFlow  based control  plane.  They demonstrate  this  approach  using  a prototype system for MPLS Traffic Engineering. Additionally, they discuss deficiencies of the MPLS  control  plane focusing  on  MPLS-TE  and suggest how a few new control applications on the network OS can be used to replace all MPLS control plane functionalities like distributed signaling and routing.
In \cite{HuiHybnet}, Hui et al. describe their experience in the design of HybNET which is a framework for automated network management of hybrid network infrastructure (both SDN and legacy network infrastructure). They discuss some of the challenges they encountered, and provide a best-effort solution in providing compatibility between legacy and SDN switches while retaining some of the advantages and flexibility of SDN enabled switches.

\subsection{Traffic  engineering  and  power management}
In some related works, the authors focus on combination of traffic engineering and power management in MPLS/SDN hybrid networks~\cite{katovHybridSDN,tajiki2017joint,GuoSDNOSPF,tajiki2018energy}. The authors of \cite{katovHybridSDN} propose a methodology for resource consolidation towards minimizing the power consumption in a large network, with a substantial resource over provisioning. The focus is on the operation of the core MPLS networks. The proposed approach is based on a SDN scheme with a reconfigurable centralized controller, which turns off certain network elements.

Some other works, explore the traffic engineering in a SDN/OSPF hybrid network. As an example, the authors of \cite{GuoSDNOSPF} propose a scenario in which the OSPF weights and flows plitting ratio of the SDN nodes can change. The controller can arbitrarily split the flows coming into the SDN nodes. The regular nodes still run OSPF. The proposed algorithm is called SOTE that can obtain a lower maximum link utilization in compared with pure OSPF networks. 

\subsection{Incremental  deployment}
Caria et al. \cite{CariaDivid}, propose a method of hybrid SDN/OSPF operation. Their method is different from other hybrid approaches, as it uses SDN nodes to partition an OSPF domain into sub-domains thereby achieving the traffic engineering capabilities comparable to full SDN operation. They place SDN-enabled routers as subdomain border nodes, while the operation of the OSPF protocol continues unaffected. In this way, the SDN controller can tune routing protocol updates for traffic engineering purposes before they are flooded into sub-domains. While local routing inside sub-domains remains stable at all times, inter-sub-domain routes can be optimized by determining the routes in each traversed sub-domain. The authors of \cite{VissicchioSafeUpdate} propose an algorithm for safely update of hybrid SDN networks.

A system for incremental deployment of hybrid SDN networks consisting of both legacy forwarding devices and programmable SDN switches is presented in \cite{HongIncr}. They propose an algorithm to determine which legacy devices to upgrade to SDN and how legacy and SDN devices can interoperate in a hybrid environment to satisfy a variety of traffic engineering (TE) goals such as load balancing and fast failure recovery.

\begin{table*}
\small
\renewcommand{\arraystretch}{1.3}
\caption{Comparison of Different Network Architecture for WAN Networks (*:~bad, ~**:~medium, ~***:~good)}
\label{Cmp_PSDN_Hyb_MPLS}
\centering
\begin{tabular}{|p{6.5cm} || p{1.5cm} | p{0.5cm} || p{3cm} | p{0.5cm} || p{1.5cm} | p{0.5cm} |}\hline
\textbf{Parameter} & \multicolumn{2}{l||}{\textbf{Pure SDN}} & \multicolumn{2}{l||}{\textbf{Hybrid Network}} & \multicolumn{2}{l|}{\textbf{Traditional MPLS}}\\ \hline\hline
\textbf{Flexibility }& high & \textbf{***} & high & \textbf{***} & medium &\textbf{**}\\ \hline
\textbf{Granularity of resource allocation} & flow-level & \textbf{***}& [flow, LSP]-level &\textbf{***}& LSP-level &\textbf{**}\\\hline
\textbf{Computational Complexity} & high &\textbf{*}& medium &\textbf{**}& medium &\textbf{**}\\\hline
\textbf{Cost of applying to the current networks} & high &\textbf{*}& low &\textbf{**}& no cost &\textbf{***}\\\hline
\textbf{Configuration} & easy & \textbf{***}& medium &\textbf{**}& hard &\textbf{*}\\\hline\hline
\multicolumn{7}{|l|}{Evaluation~~~~~ *:~bad,~~~~~~~~~ **: medium,~~~~~~~~~ ***:~good.}\\\hline
\end{tabular}\\
\end{table*}

\subsection{Novelty and Comparison}
The most important drawbacks of the existing algorithms are categorized into two main classes: a) fixed allocation of resources to the flows and b) do not considering the impact of flows on each other. In order to explain the impact of fixed allocation of resources to the flows, consider flow $x$ is routed via path $y$. In most of the existing algorithms, the flow continues streaming from this path even if it reduces/increases its rate by multiple order of magnitude. This results in congestion or low link utilization.

\begin{figure}
\centering
\includegraphics[scale=0.65]{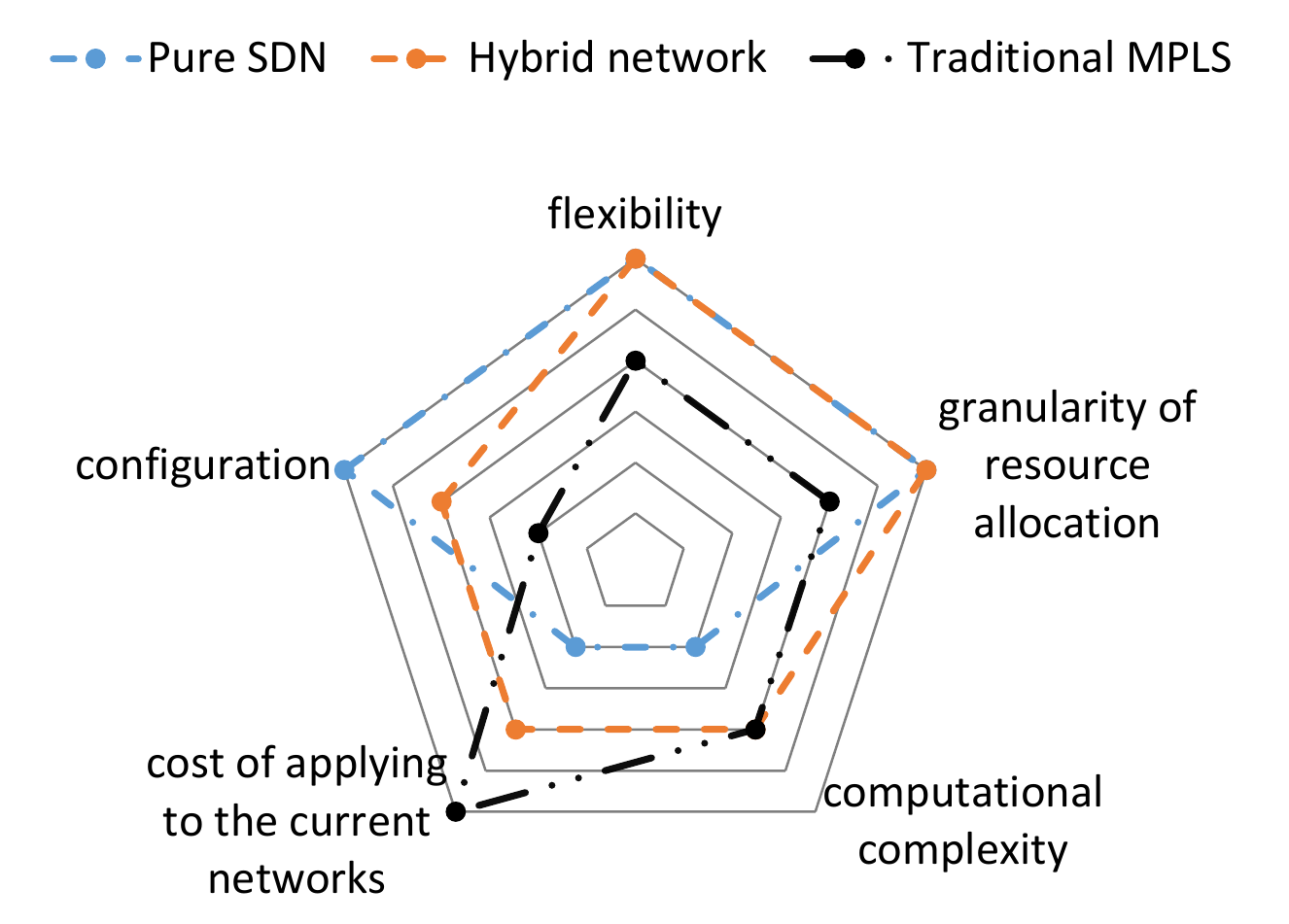}
\caption{Comparison of Different Network Architecture for WAN Networks.}
\label{SpiderChart_SDN_Hybrid_MPLS_Comparison}
\end{figure}

In order to manage or upgrade the MPLS networks, there are three main architectures: 1- pure SDN (all switches are OpenFLow-enabled) 2- hybrid (OpenFlow-enabled and conventional MPLS routers) 3- pure MPLS (conventional MPLS routers). In Table \ref{Cmp_PSDN_Hyb_MPLS}, these three architectures are compared from different measurements. In order to simplify the process of understanding the differences, Fig. \ref{SpiderChart_SDN_Hybrid_MPLS_Comparison} is depicted. As can be seen, hybrid networks provide a trade-off between different metrics while they are applicable for current MPLS networks.

The major differences of our work with the traditional approaches are as follows: 1) lots of traditional approaches focus on the routing of new flows while our approach (STEM: SDN-based Traffic Engineering in MPLS) focuses on the re-routing of existing flows and re-creation of LSPs \footnote{An LSP is a predetermined path from a source router to a destination router}. 2) since STEM considers the effect of flows on each other, it can handle the problem of resource partitioning. 3) despite the traditional algorithms, STEM can be used along with any other routing algorithm. 4) STEM focuses on network reconfiguration overhead and re-routes the flows in a way that minimizes the network reconfiguration overhead 5) STEM adds the flexibility of SDN-based approaches to existing MPLS networks by adding a few number of low-cost OpenFlow-enabled switches to them. 

\section{The Proposed Architecture}\label{ProposedArchitecture}
    In this section, problem definition and a quick overview of the proposed architecture is presented, thereafter, the comprehensive details of the proposed architecture components is discussed.
    \subsection{Problem Definition}\label{problemDefinition}
        The considered network consists of three main parts 1) MPLS routers as the core of the network, 2) low-cost OpenFlow-enabled switches as the edge of the network, and 3) a central controller such as ONOS \cite{berde2014onos}. All of the MPLS routers and OpenFlow switches are configurable via PCEP and OpenFlow protocols, respectively. Since the Edge switches are all OpenFlow-enabled, the protocol used for communication of these switches and the controller is OpenFlow. Therefore, the controller can query the switches for this part of the network topology and traffic matrix. On the other hand, since the core network runs MPLS, the controller should support PCEP protocol (ONOS controller has a PCE element). PCE element is the component which is responsible for communicating with the MPLS routers via PCEP protocol and assigning the LSP to the links. The controller can gather information from the MPLS routers via querying them, too.
        
        The problem is to find a novel traffic engineering architecture and routing/re-routing algorithm in which the integration of OpenFlow and traditional MPLS is adopted, i.e., proposing an architecture where SDN is going to be deployed in an existing network. The objective is to propose a traffic engineering scheme for integration of MPLS and OpenFlow networks that can adaptively and dynamically manage traffic in a network to accommodate different traffic patterns.
        
        \begin{figure*}
        \includegraphics[scale=0.5]{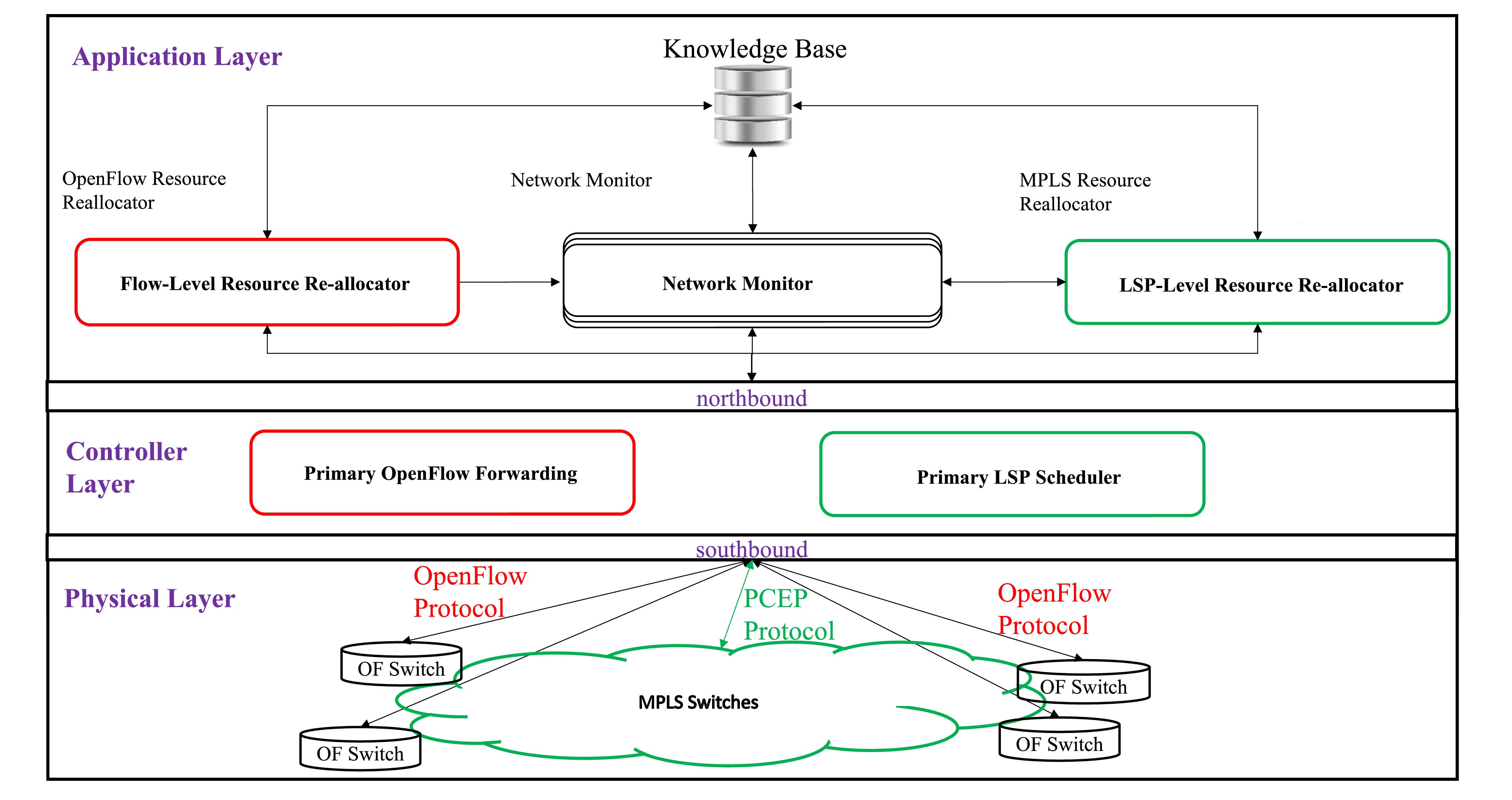}
        \caption{The Proposed Network Architecture.}
        \label{fig:ProposedArchitecture}
        \end{figure*}
    \subsection{Overview of the Proposed Architecture}
    In this subsection, a brief overview of the proposed architecture and its components is presented. We assume that there is a centralized SDN controller computing the forwarding table for the OF switches as well as providing LSP for MPLS routers. To this end, the controller peers with the network and gathers information about the network traffic and topology. The OF switches along with the forwarding of the packets, do some simple traffic measurement and forward these measurement to the controller.  In order to dynamically adapt the network configuration with respect to the traffic variations, the controller exploits this traffic information along with information gathered from the MPLS network to update  these tables at the OF switches. It is notable that LSP reconfiguration is mandatory when flow re-routing (in OF switches) is not sufficient for congestion control. It is also worth noting that in our proposal we jointly use the existing protocols along with a new rescheduling algorithm. As can be seen in Fig. \ref{fig:ProposedArchitecture}, the proposed architecture has the following modules: 
    \begin{itemize}
    \item Primary OpenFlow forwarding: a common routing algorithm which runs when there is a new arrival flow, e.g., ECMP protocol.
    \item Flow-level resource re-allocator: an algorithm runs when there is a network congestion or the predefined time interval is elapsed. 
    \item Primary LSP scheduler: an existing LSP scheduler, e.g., RSVP-TE protocol. 
    \item LSP-level resource re-allocator: an algorithm which runs when the “flow-level resource re-allocator” cannot handle the current network traffic using the existing LSPs and requests a LSP rescheduling.
    \item Network monitoring: periodically monitors the links’ state, updates the “knowledge base”, and provides traffic matrix for the “flow-level resource re-allocator” module.
    \end{itemize}
    \begin{table*}[!t]
    \small
    \renewcommand{\arraystretch}{1.3}
    \caption{Symbols Definitions}
    \label{Symbols}
    \centering
    \begin{tabular}{|p{1cm} | p{6.95cm} || p{1cm} | p{7cm} |}\hline
    \textbf{Symbol} & \textbf{Definition} & \textbf{Symbol} & \textbf{Definition}\\ \hline\hline
    $\mu$ & Maximum link utilization & $N_S$ & Number of switches\\ \hline 
    $N_{L}$ & Number of LSPs &  $N_{F}$ & Number of flows  \\ \hline 
    $\mathbf{s}_L$ & a $1 \times N_L$ vector denoting the start nodes of LSPs & $\mathbf{B}$ & a $N_S\times N_S$ matrix denoting the Links bandwidth\\ \hline 
    $\mathbf{e}_L$ & a $1 \times N_L$ vector denoting the end (destination) nodes of LSPs & $\mathbf{td}_F$ & a $1\times N_F$ vector denoting the maximum tolerable delay of flows \\ \hline 
    $\mathbf{pd}_L$ & a $1\times N_F$ vector denoting the propagation delay of LSPs & $\mathbf{D}$ & a $N_S\times N_S$ matrix denoting the propagation delay of links\\ \hline
    $\mathbf{c}_L$ & a $1\times N_L$ vector denoting the capacity of LSPs & $\mathbf{s}_F$ & a $1\times N_F$ vector denoting the start node of flows\\ \hline
    $\mathbf{r}_F$ & a $1\times N_F$ vector denoting the rate of flows & $\mathbf{e}_F$ & a $1\times N_F$ vector denoting the end node of flows \\ \hline
    \multicolumn{2}{|l||}{\textbf{Problem Variable (binary)}} & \multicolumn{2}{l|}{\textbf{Problem Variable (binary)}}\\ \hline
    $\mathbf{FR}$ & a $N_F\times N_L$ matrix denoting the assignment of flows to the LSPs & $\mathbf{LR}$ & a $N_S\times N_S\times N_L$ matrix denoting the assignment of links to LSPs \\ \hline
    \end{tabular}\\
    \end{table*}
    
    \subsection{Comprehensive Discussion of the Proposed Architecture}
    In this subsection, each component of the proposed architecture is precisely discussed. It should be mentioned that the \textit{Knowledge Base} element is used to gather information about the previous states of the network to predict the future state of the network\footnote{Network monitoring and traffic prediction algorithms are out of the scope of this work and we consider these elements are designed perfectly.}.
    \subsubsection{Primary OpenFlow Forwarding Element}
    selects an appropriate LSP for the new flows. This component works based on the existing algorithms such as shortest-path or ECMP. Therefore, it is a traditional routing algorithm (not a re-routing algorithm) and it does not consider the impact of flows on each other. This element should be implemented as a part of the controller to enhance the performance of the routing scheme.
    \subsubsection{Primary LSP Scheduler Element}
    A Path Computation Element (PCE\cite{RFC5440}) is an entity that can compute a path based on a network graph. A Path Computation Client (PCC) is any client application requesting from PCE to compute a path. The Path Computation Element Protocol (PCEP) enables communications between between two PCEs or a PCE and a PCC. \textit{Primary LSP Scheduler} is a PCE. If a new LSP is required, this component is invoked to create a new LSP. Current controllers such as ONOS\cite{OnosController} and OpenDayLight\cite{OpenDaylightController} support PCEP. Since \textit{Primary OpenFlow Forwarding} works based on the existing protocols, it does not consider the impact of LSPs on each other. This element should be implemented as a part of the controller to enhance the performance of the routing scheme.
    \subsubsection{Flow-Level Resource Re-Allocator Element}
    The most important role of OpenFlow switches in our proposed architecture is the assignment of flows to the existing LSPs. This element is designed to control the network congestion. In order to avoid congestion in the links, "flow-level resource re-allocator element" re-routes some of the flows when the maximum link utilization exceeds a predefined threshold. At this time, it re-assigns flows to the existing LSPs. To this end, we mathematically formulate an optimization problem at which the main aim is to control the traffic congestion by re-assigning the flows to the LSPs subject to the flow tolerable delay, the flow conservation constraint and LSP bandwidth restriction. Besides, the proposed optimization problem minimizes the reconfiguration overhead. 
    \subsubsection{LSP-Level Resource Re-Allocator Element(LR)}
    The network side-effect of "flow-level resource re-allocator" is sufficiently lower than the side-effect of this element. Therefore, just in the cases that the "flow-level resource re-allocator" could not control the network congestion, it sends an LSP-reassignment request to "LSP-level resource re-allocator". The LSP re-allocator element re-assigns links to the LSPs to reduce the traffic load of the congested links. To this end, we mathematically formulate an optimization problem at which the main aim is to route requested LSPs subject to the link capacity restriction, LSP conservation constraints, and requested end-to-end propagation delay restriction of LSPs. Besides, the corresponding optimization problem minimizes the network changes to reduce the side-effect of network re-configuration. Since this optimization problem is in form of binary linear programming, we can adopt the well-known and efficient branch and cut method to obtain an optimal solution. 

\section{Problem Formulation}\label{problemForumlation}
    In order to implement the proposed architecture, two main tasks must be done: 1) re-routing of networks flows (re-assignment of flows to the LSPs) 2) re-creation of LSPs based on the network dynamics. To do these tasks, we mathematically formulate these optimization problems in this section. Table \ref{Symbols} contains all the symbols which are used in the formulations. The variables $N_L$, $N_S$, and $N_F$ specify the number of LSPs, switches, and flows, respectively while $\mathbf{c_L}$, $B$, and $r_F$ represent the LSP capacity, link bandwidth, and flow rate, respectively. The vectors $(\mathbf{s_L}, \mathbf{e_L})$ and $(\mathbf{s_F}, \mathbf{e_F})$ represent the (source, destination) of LSPs and flows, respectively. For each LSP and link, $\mathbf{pd_L}$ and $D$ specify the propagation delay, respectively while $\mathbf{td_F}$ specifies the maximum tolerable delay of flows. The assignment of flows to the LSPs is presented using the matrix $\mathbf{FR}$. Finally, Matrix $\mathbf{LR}$ denotes the assignment of links to the LSPs.

    \subsection{Flow Re-Routing}
    In the following, the problem of assigning the ingress flows to the LSPs in a way that minimizes the network reconfiguration overheads is presented. The problem  formulation is in form of Binary Linear Programming (BLP).  
    \begin{subequations}\label{FlowSelectionEq}
    \begin{align}
        &\min_{\mathbf{FR}}(|{\mathbf{FR}}-{\mathbf{FR}}_{old}|),\label{FlowSelectionEqObj}\\&
        \nonumber\text{Subject to: } \\&
        \hspace{8px}\sum_{f=1}^{N_F}{(\mathbf{r}_F[f]\times \mathbf{FR}[f,i])}\leq \mathbf{c}_L[i],\forall i\in \{1,...,N_L\},\label{FlowSelectionEqOne}\\&
        \hspace{8px}\sum_{i=1}^{N_L}{(\mathbf{FR}[f,i]\times \mathbf{pd}_L[i])}\leq \mathbf{td}_F[f],\forall f\in \{1,...,N_F\},\label{FlowSelectionEqTwo}\\&
        \label{FlowSelectionEqThree}\hspace{8px} \sum_{i=1}^{N_L}{\mathbf{FR}[f,i]}=1, \forall f\in \{1,...,N_F\},\\&
        \hspace{8px} \mathbf{FR}[f,i]\times \mathbf{s}_F[f]=\mathbf{FR}[f,i]\times \mathbf{s}_L[i], \nonumber\\&
            \label{FlowSelectionEqFour}\hspace{16px} \forall f\in \{1,...,N_F\}, \forall i\in \{1,...,N_L\},\\&
        \hspace{8px}\mathbf{FR}[f,i]\times \mathbf{e}_F[f]=\mathbf{FR}[f,i]\times \mathbf{e}_L[i],\nonumber\\&
            \label{FlowSelectionEqFive}\hspace{16px}\forall f\in \{1,...,N_F\}, \forall i\in \{1,...,N_L\},\\&
        \hspace{8px}\mathbf{FR}[f,i]\in \{0,1\}, \forall f \in \{1,...,N_F\},\nonumber\\&
            \label{FlowSelectionEqSix}\hspace{16px}\forall i \in \{1,...,N_L\}.
    \end{align}
    \end{subequations}
    where the objective function \eqref{FlowSelectionEqObj} minimizes the reconfiguration overhead by reducing the number of flows that are changed. Eq. \eqref{FlowSelectionEqOne} guarantees the rate of flows on each LSP to be less than the LSP's capacity. Eq. \eqref{FlowSelectionEqTwo} seeks for the propagation delay of the selected LSP and compares it with the tolerable delay of the flows. Since each flow must assign to one and only one LSP, Eq. \eqref{FlowSelectionEqThree} is considered as a part of this optimization problem. Equations \eqref{FlowSelectionEqFour} and \eqref{FlowSelectionEqFive} ensure that the start and end points of the selected LSP is similar to the start and end of the corresponding flow.
    
    If the required resources of all LSPs are reserved in the MPLS routers (e.g., using RSVP-TE protocol) then the optimization problem \eqref{FlowSelectionEq} is used to re-assign flows to the LSPs. However, if there is one or more LSPs that do not reserve the required resources then the optimization problem should be formulated as follows:
    \begin{subequations}\label{NoResrvRescEq}
    \begin{align}
        &\min_{\mathbf{FR}}(|{\mathbf{FR}}-{\mathbf{FR}}_{old}|),\\&
        \nonumber\text{Subject to: } \\&
        \hspace{8px}\nonumber\eqref{FlowSelectionEqOne}, \eqref{FlowSelectionEqTwo}, \eqref{FlowSelectionEqThree}, \eqref{FlowSelectionEqFour}, \eqref{FlowSelectionEqFive}, \eqref{FlowSelectionEqSix}\\&
        \nonumber\hspace{8px}\sum_{f=1}^{N_F}{\sum_{i=1}^{N_L}{(\mathbf{r}_F[f]\times \mathbf{FR}[f,i]\times \mathbf{LR}[i,j,v])}}\leq \mu \mathbf{B}[j,v],\\
            &\hspace{16px}\forall j,v\in \{1,...,N_S\}.
    \end{align}
    \end{subequations}
    
    \subsection{LSP Re-Creation}
\begin{figure}[!htbp]
        \centering
        \begin{subfigure}{0.45\columnwidth}
          \centering
          \includegraphics[width=1\linewidth]{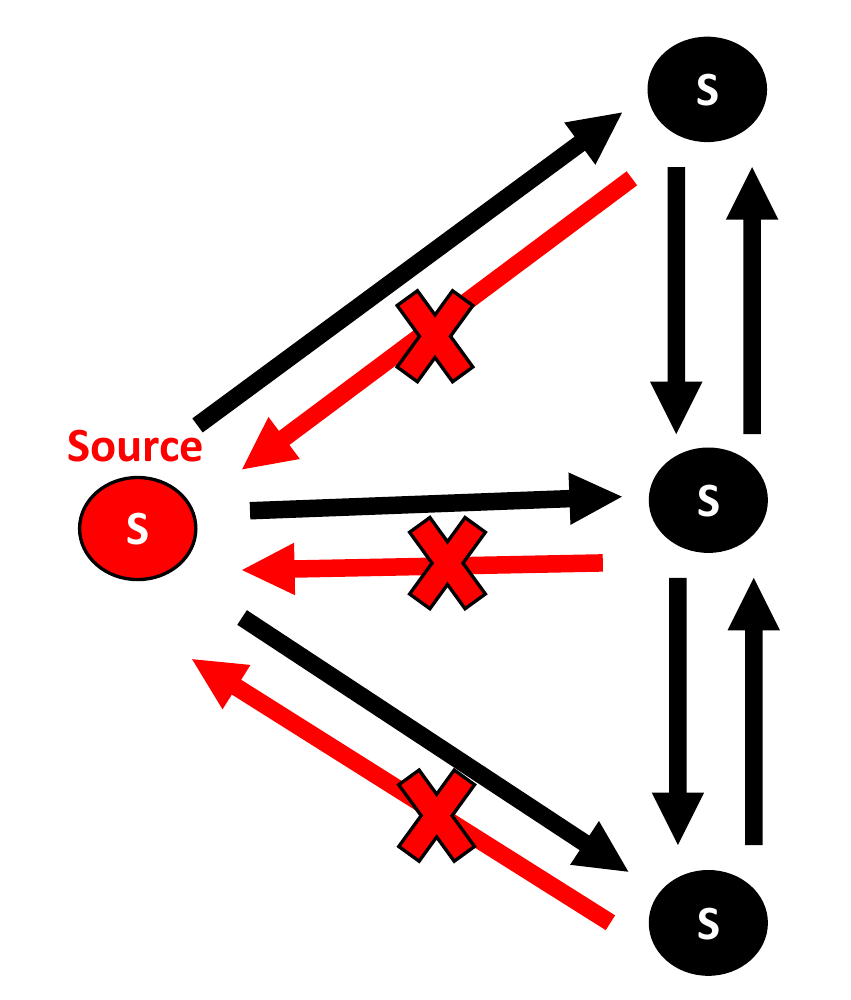}
          \caption{First Part of  Eq. \eqref{LSPCreationEqThree}}
          \label{fig:SrcConstrnt}
        \end{subfigure}
        \begin{subfigure}{0.45\columnwidth}
          \centering
          \includegraphics[width=1\linewidth]{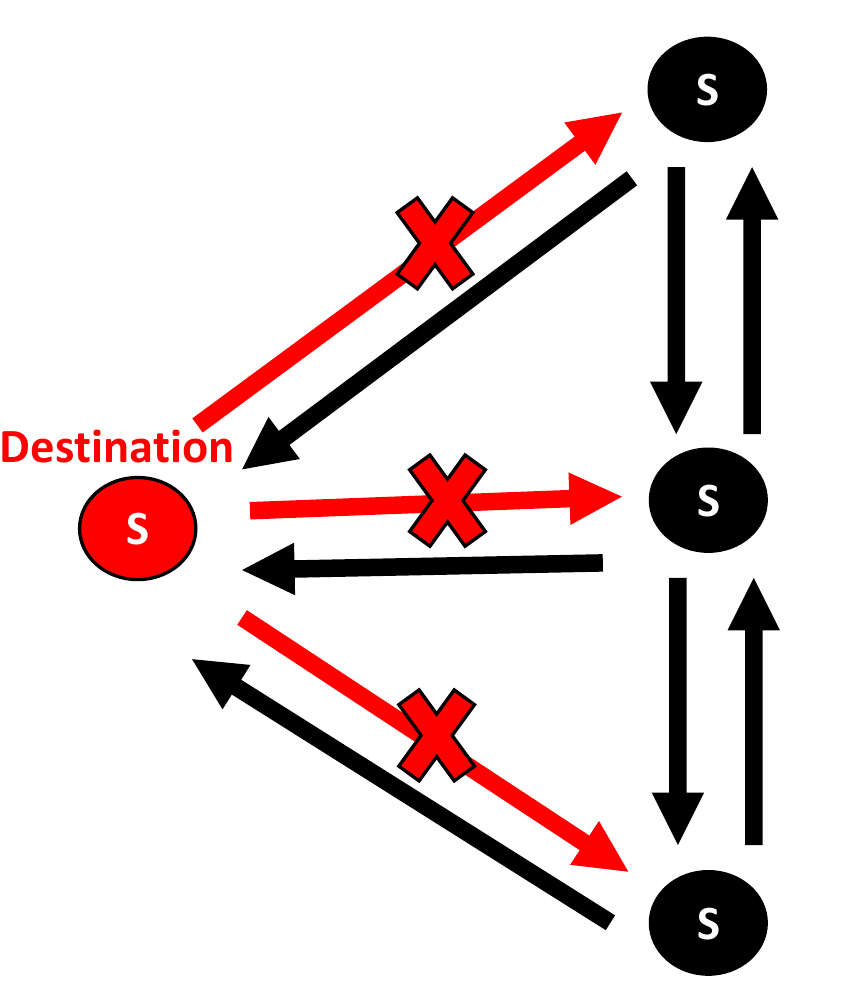}
          \caption{Second Part of Eq. \eqref{LSPCreationEqThree}}
          \label{fig:DestCnstrnt}
        \end{subfigure}
        \begin{subfigure}{0.3\textwidth}
          \centering
          \includegraphics[width=0.8\linewidth]{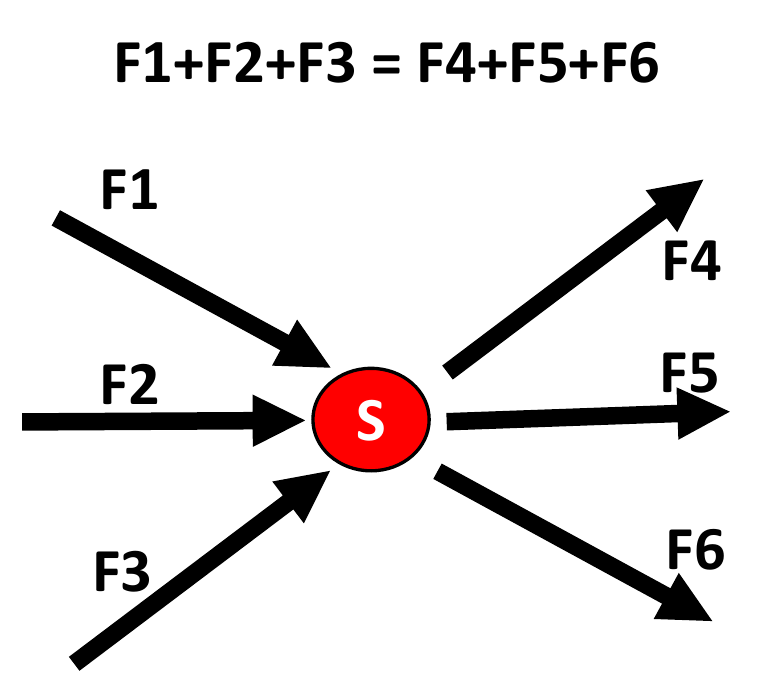}
          \caption{Eq. \eqref{LSPCreationEqSix}}
          \label{EEEConst}
        \end{subfigure}
        \caption{Visual Illustration of Constraints.}
        \label{fig:VisualIllustration}
    \end{figure}

    Each LSP is a sequence of links from a specified source to a specified destination. Since the re-creation of the LSPs (re-assignment of links to the LSPs) has effect on the ongoing traffics, we try to concur the traffic dynamic nature via flow re-routing instead of changing the LSPs. However, if the flow re-routing could not handle this dynamicity with the current LSPs then the LSPs should be re-created. In this way, we mathematically formulate the problem of LSP re-creation and explore a solution to solve the corresponding optimization problem. We extend our previous work \cite{QNR} to match this problem. The formulation is in form of binary linear programming as follows:
    
    \begin{subequations}\label{LSPCreationEq}
    \begin{align}
    \label{LSPCreationEqObj}&\min_{\mathbf{LR}}{(|\mathbf{LR}-\mathbf{LR}_{old}|)},\\
    &\nonumber\text{Subject to:}\\
    \label{LSPCreationEqOne}&\hspace{8px}\sum_{i=1}^{N_L}{(\mathbf{LR}^{i}_{j,v}\times \mathbf{c}_L[i])}\leq \mu \mathbf{B}_{j,v}, \forall j,v\in \{1,...,N_S\},\\
    \label{LSPCreationEqTwo}&\hspace{8px}\sum_{j=1}^{N_S}{\sum_{v=1}^{N_S}{({\mathbf{LR}^{i}_{j,v}}\times \mathbf{D}[j,v])}}\leq \mathbf{pd}_L[i], \forall i\in \{1,...,N_L\},\\
    \label{LSPCreationEqThree}&\hspace{8px}\sum_{j=1}^n{\mathbf{LR}^{i}_{j,\mathbf{s}_L[i]}}= \sum_{v=1}^n{\mathbf{LR}^{i}_{\mathbf{e}_L[i],j}}=0, \forall i \in \{1,..., N_L\},\\
    \label{LSPCreationEqFour}&\hspace{8px}\sum_{j=1}^n{\mathbf{LR}^{i}_{\mathbf{s}_L[i],j}}= \sum_{j=1}^n{\mathbf{LR}^{i}_{j,\mathbf{e}_L[i]}}=1, \forall i \in \{1,..., N_L\},\\
    &\hspace{8px}\sum_{i=1}^p{{\mathbf{LR}^{i}_{j,v}}}=\sum_{i=1}^p{{\mathbf{LR}^{i}_{v,j}}}, \forall i\in \{1,...,N_L\}, \nonumber\\&
        \hspace{16px}\forall j\in \{1,...,N_S\}-\{\mathbf{s}_L[i],\mathbf{e}_L[i]\},\label{LSPCreationEqFive}\\
    \label{LSPCreationEqSix}&\hspace{8px}\sum_{v=1}^n{\mathbf{LR}^{f}_{j,v}}\leq 1, \forall j\in \{1,...,N_S\}, \forall i \in \{1,...,N_L\},\\
    \label{LSPCreationEqNine}&\hspace{8px}\mathbf{LR}^{i}_{j,v}\in \{0,1\}, \forall i\in \{1,...,N_L\}, \forall j,v \in \{1,...,N_S\}.
    \end{align}
    \end{subequations}
    where the objective function \eqref{LSPCreationEqObj} minimizes the reconfiguration overhead by reducing the number of LSPs that are changed. The Eq. \eqref{LSPCreationEqOne} guarantees the rate of flows on each link to be less than the link's capacity. Eq. \eqref{LSPCreationEqTwo} seeks for the propagation delay of the selected path and compares it with the tolerable delay of the requested LSP. Fig. \ref{fig:SrcConstrnt} and \ref{fig:DestCnstrnt} illustrate the Eq. \eqref{LSPCreationEqThree} where the streams are enforced to leave the source switches and enter to the destination one.

\begin{figure}[!htbp]
        \centering
        \begin{subfigure}{1\columnwidth}
          \centering
          \includegraphics[width=1\linewidth]{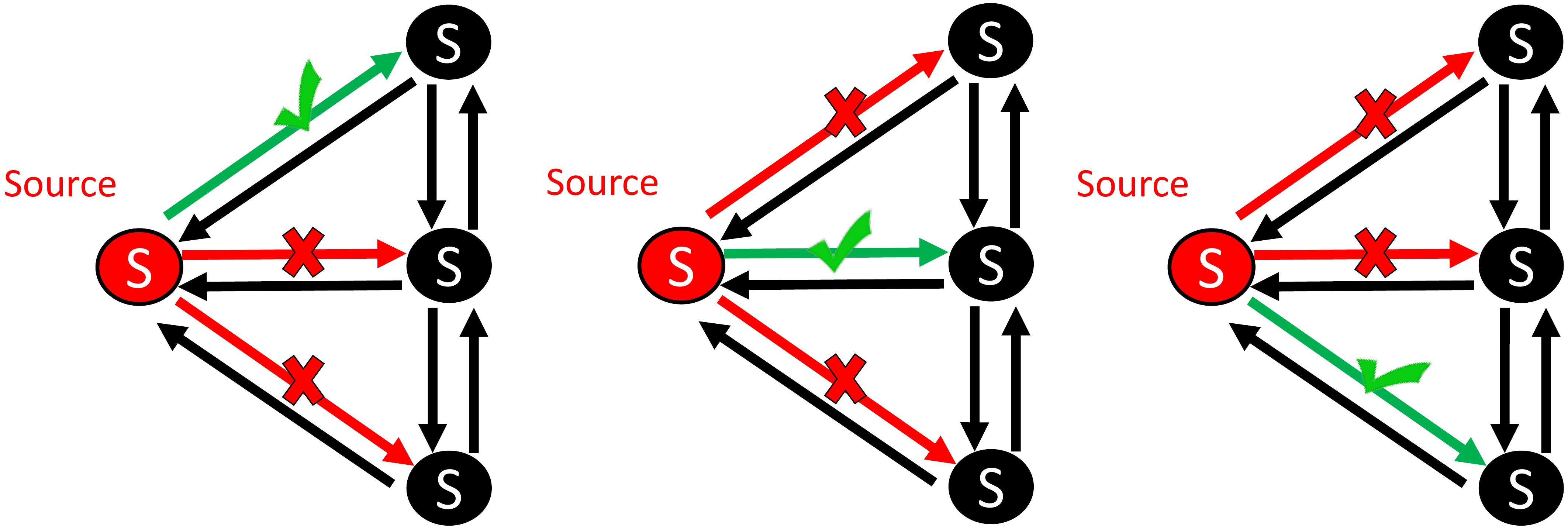}
          \caption{First Part of Eq. \eqref{LSPCreationEqFour}}
          \label{fig:LeaveSrcEntrDstConst}
        \end{subfigure}
        \begin{subfigure}{1\columnwidth}
          \centering
          \includegraphics[width=1\linewidth]{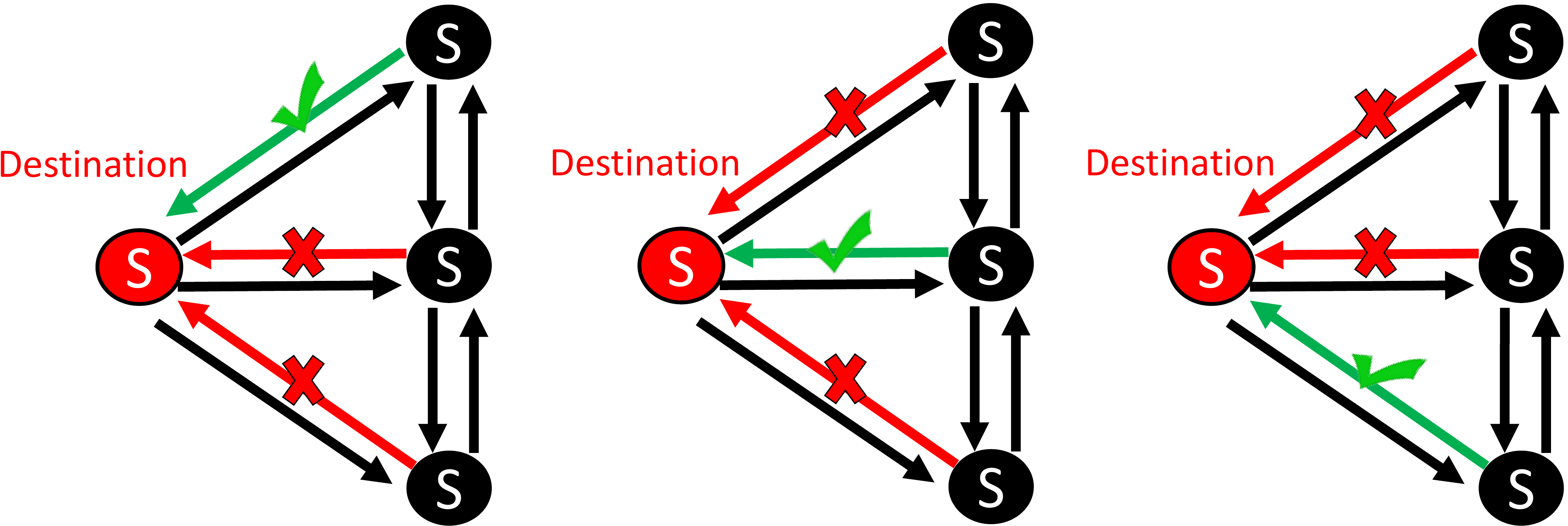}
          \caption{Second Part of Eq. \eqref{LSPCreationEqFour}}
          \label{fig:EntrDstConstrnt}
        \end{subfigure}
        \caption{Visual Illustration of Constraints.}
        \label{fig:VisualIllustration2}
    \end{figure}

    It should be mentioned that the stream cannot return to the source switch or leaves the destination one. To this end, Eq. \eqref{LSPCreationEqFour} is included in this formulation. Fig. \ref{fig:LeaveSrcEntrDstConst} and \ref{fig:EntrDstConstrnt} visually illustrate the mentioned constraint. To prevent loops in the selected paths, Eq. \eqref{LSPCreationEqSix} is considered which is depicted in Fig. \ref{EEEConst}.
    
\section{Fast Flow Re-routing Heuristic (FFR)}\label{sec:hueristic}
    \begin{algorithm}[!bp]
    	\caption{Fast flow re-routing heuristic}
    	\label{alg:FFR}
    	\normalsize
    	\allowdisplaybreaks
    	\begin{algorithmic}[1]
        	\break
        	\INPUT{Set of flows, Set of LSPs}
        	\OUTPUT{Assignement of flows to LSPs}
        	\For{each flow $f$ in $\mathcal{F}$}
            	\State{$L=Find\_Proper\_LSPs()$}
            	\For{each LSP $lsp$ in $\mathcal{L}$}
            	    \If{$Free\_Capacity(lsp)>=Size(f)$}
            	        \State{set $flag=true$}
            	        \State{break}
            	    \EndIf
            	\EndFor
            	\If{not flag}
            	    \For{each LSP $lsp$ in $\mathcal{L}$}
            	        \If{$Check\_Congestion(lsp, flow)$}
            	            \State{set $flag=true$}
            	            \State{break}
            	        \EndIf
            	    \EndFor
            	\EndIf
            	\If{$flag$}
            	    \State{assign $lsp$ to $f$}
            	    \State{reduce $lsp$ size}
            	\Else
            	    \State{$LSP\_Recreation()$}
            	\EndIf
        	\EndFor\\
    	    \Return{assignments}
    	\end{algorithmic}
	\end{algorithm}
	
    Since the process of flow re-routing should be done in a real-time manner, we propose a heuristic algorithm called Fast Flow Re-routing (FFR) which is presented in  Algorithm \ref{alg:FFR}. FFR re-routs one flow in each step (line 1 of the algorithm), however, it considers the impact of previously re-routed flows on the other flows. In other words, when FFR re-routes a flow, it reduces the free capacity of the newly selected LSP. To this end, for each flow it finds all LSPs that have a similar source and destination with the flow and puts them in variable $L$ (line 2). After that, in lines 3-8, FRR probes among the $L$'s elements to find an LSP which has a free capacity more thank the flow size. If such LSP is found then variable $flag$ would set to $true$.
    
    Sequential assignment of resources may cause resource partitioning. To cope this, if the variable $flag$ is not set to $true$ (line 9) then FFR tries to find a proper LSP by adding the free capacity of links to the LSP and comparing the new LSP capacity with the flow size (lines~10-16). In lines 17-19, the selected LSP is assigned to the flow, however, if a proper LSP is not found then the $LSP\_Recreation$ function would invoked. 
\subsection{Computational Complexity}
    In this part we calculate the worst case for computational complexity of FFR. The computational complexity of lines~1, 3, and 10 are $N_F$, $N_L$, and $N_L$, respectively. The computational complexity of $Free\_Capacity$ is $N_L$ since it should search among all LSPs to find those that are proper for the flow. On the other, since each path is consist of at most $N_L$ hops then $Check\_Congestion$ is in order of $N_S$. The computational complexity of $LSP\_Recreation$ is highly dependent on the implementation approach (e.g., reference~\cite{tajiki2018CECT} propose a solution which is linear on the number of flow, switches, and paths); In our simulation we used CVX to solve this function. The computational complexity of the other parts are in order of $O(1)$. Considering $C_L$ as the computational complexity of the function $LSP\_Recreation$, the computational complexity of FFR is $O(N_F\times(N_L+N_L+N_L\times N_S + C_L)\approx O(N_F\times N_L\times N_S + C_L)$. 

\section{Performance Evaluation}\label{resultAnalysis}
In this section, the proposed scheme is compared with shortest path algorithm in which the cost function is the length of the path. The evaluation is performed via three different metrics: 
\begin{itemize}
\item System throughput: the sum of the data rates that are delivered to all terminals in a network. It is a measure to show the performance of the network;
\item Path length: the average number of steps along the selected paths for all flows. It is a measure of the efficiency of transport on a network;
\item Link utilization: the amount of data on the link divided by the total capacity of the link. It is a measure of protocol fairness.
\end{itemize} 
\subsection{Scenario Description}
We implement a traffic generator to test the performance of the proposed scheme over different network traffic scenarios. In the traffic generator, the average bandwidth demand of a flow is a fraction $B^f$ of the capacity of links, i.e., it is $B^f\times link\_bandwidth$. Rate of generated flows follows a uniform distribution between 0 and 2 times of the average rate of flows. Moreover, $F_s$ and $F_m$ are input parameters that control the number of generated flows per source switches. More precisely, the number of generated flows for each source switch follows a truncated geometric distribution with $1/(F_s \times \tau\times N)$ as the success probability and $F_m$ as the maximum number of flows. In our experiment we set the $B^f=0.08$, $F_m=10$, and $F_s$ has two values $\{0.6, 0.8\}$. Different traffic scenarios are presented in Table~\ref{tab:TrafficScenarios}.
\begin{table}[!htbp]
    \caption{Different Traffic Scenarios.}
    \label{tab:TrafficScenarios}
    \centering
    \begin{tabular}[t]{|l|c|c|c|c|}\hline
         & $F_s$ & $N_F$ & $F_m$ & $B^f$\\ \hline\hline
        Scenario 1 & 0.8 & 84 & 10 & 0.08\\ \hline
        Scenario 2 & 0.8 & 86 & 10 & 0.08\\ \hline
        Scenario 3 & 0.6 & 71 & 10 & 0.08\\ \hline
        Scenario 4 & 0.6 & 70 & 10 & 0.08\\ \hline
    \end{tabular}
\end{table}

\subsection{Simulation Setup}
\begin{figure}[!htbp]
    \centering
    \includegraphics[scale=0.3]{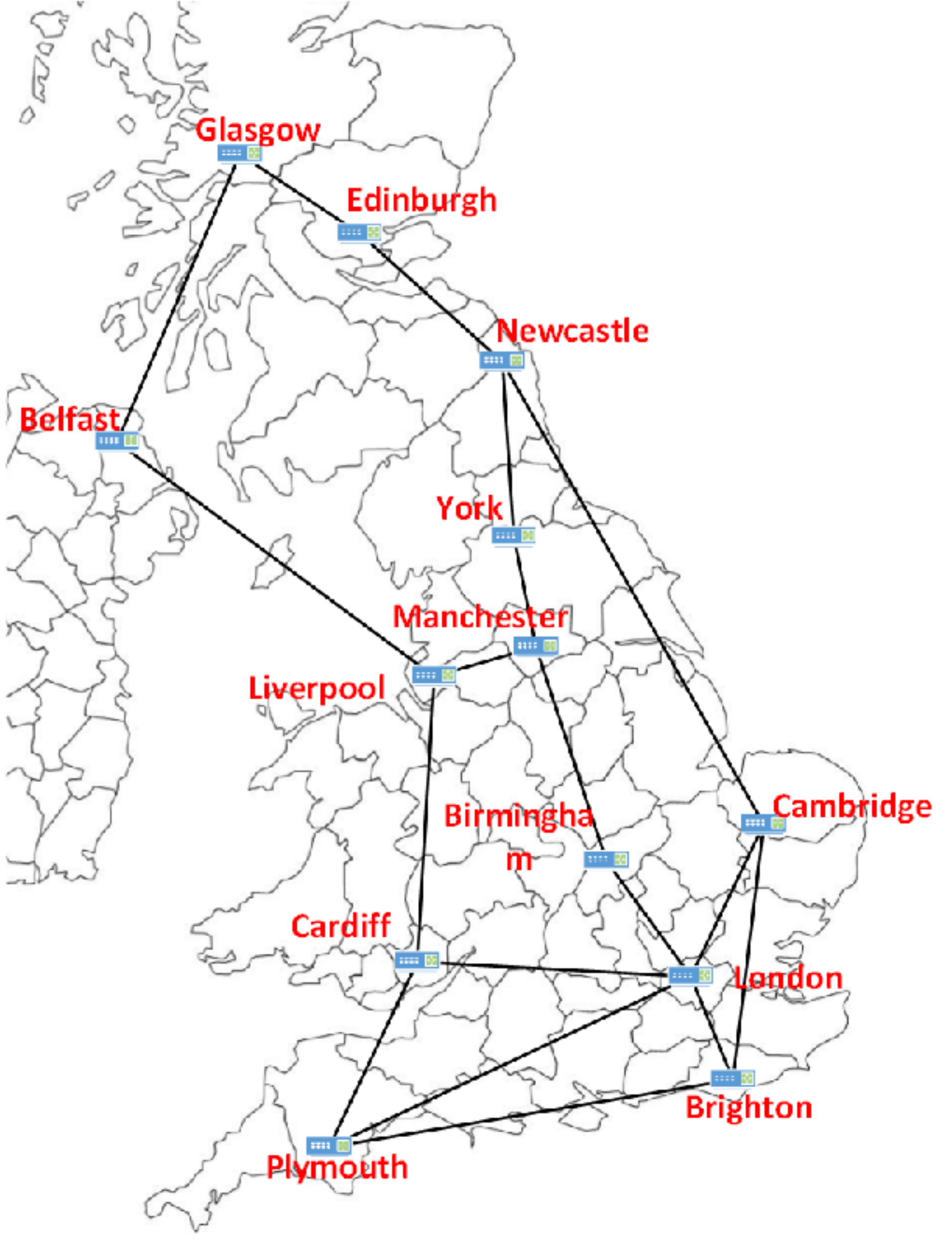}
    \caption{The Considered Topology.}
    \label{UKmapTopology}
\end{figure}

In this subsection, the network topology and traffic pattern used in our simulation is described. The topology is inspired by the work \cite{CrtngAndMngDyn} and depicted in Fig. \ref{UKmapTopology}. For the sake of simplicity, all links' propagation delay are considered equal. The simulation is done using MATLAB R2016b and the hardware configuration of the PC is represented in Table~\ref{tab:SystemConfiguration}.

\begin{table}[!htbp]
	\caption{Hardware Configuration.}
	\label{tab:SystemConfiguration}
	\rowcolors{2}{gray!25}{white}
	\resizebox{\columnwidth}{!}{
	\begin{tabular}[t]{|l|l|} 
		\hline
		\textbf{Name} & \textbf{Description}\\
		\hline\hline
        Processor & Intel(R) Core(TM) i5-2410M CPU @ 2.30GHz\\\hline
        IDE & Standard SATA AHCI Controller \\\hline
        RAM & 4.00 GB\\\hline
        System Type & 64-bit Operating System, Windows 10\\\hline
	\end{tabular}}
    \end{table}

\subsection{Throughput Results}
\begin{figure}[!htbp]
\centering
\begin{subfigure}{0.49\columnwidth}
  \centering
  \includegraphics[width=1\linewidth]{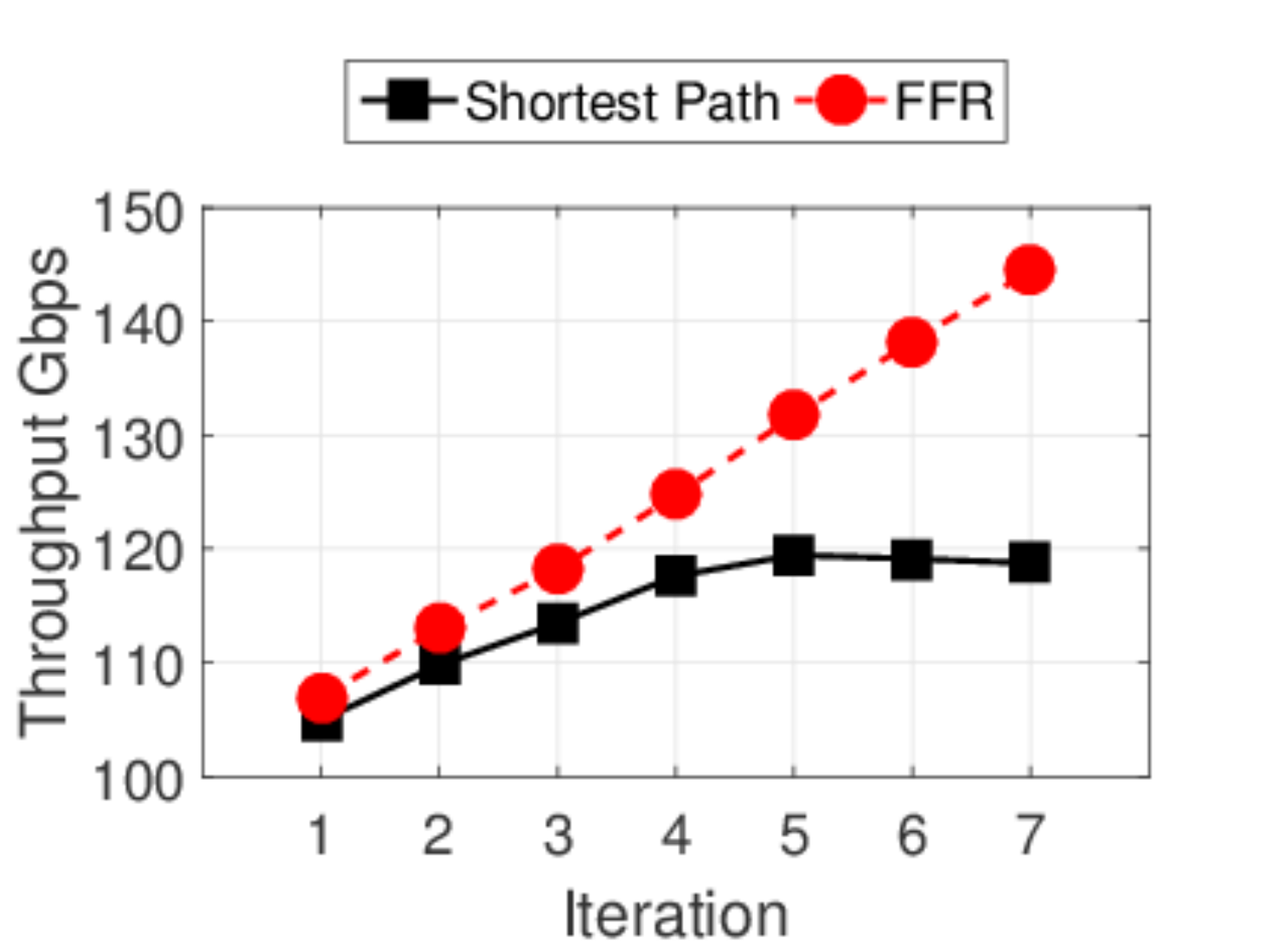}
  \caption{Scenario 1}
  \label{fig:fig84th}
\end{subfigure}
\begin{subfigure}{0.49\columnwidth}
  \centering
  \includegraphics[width=1\linewidth]{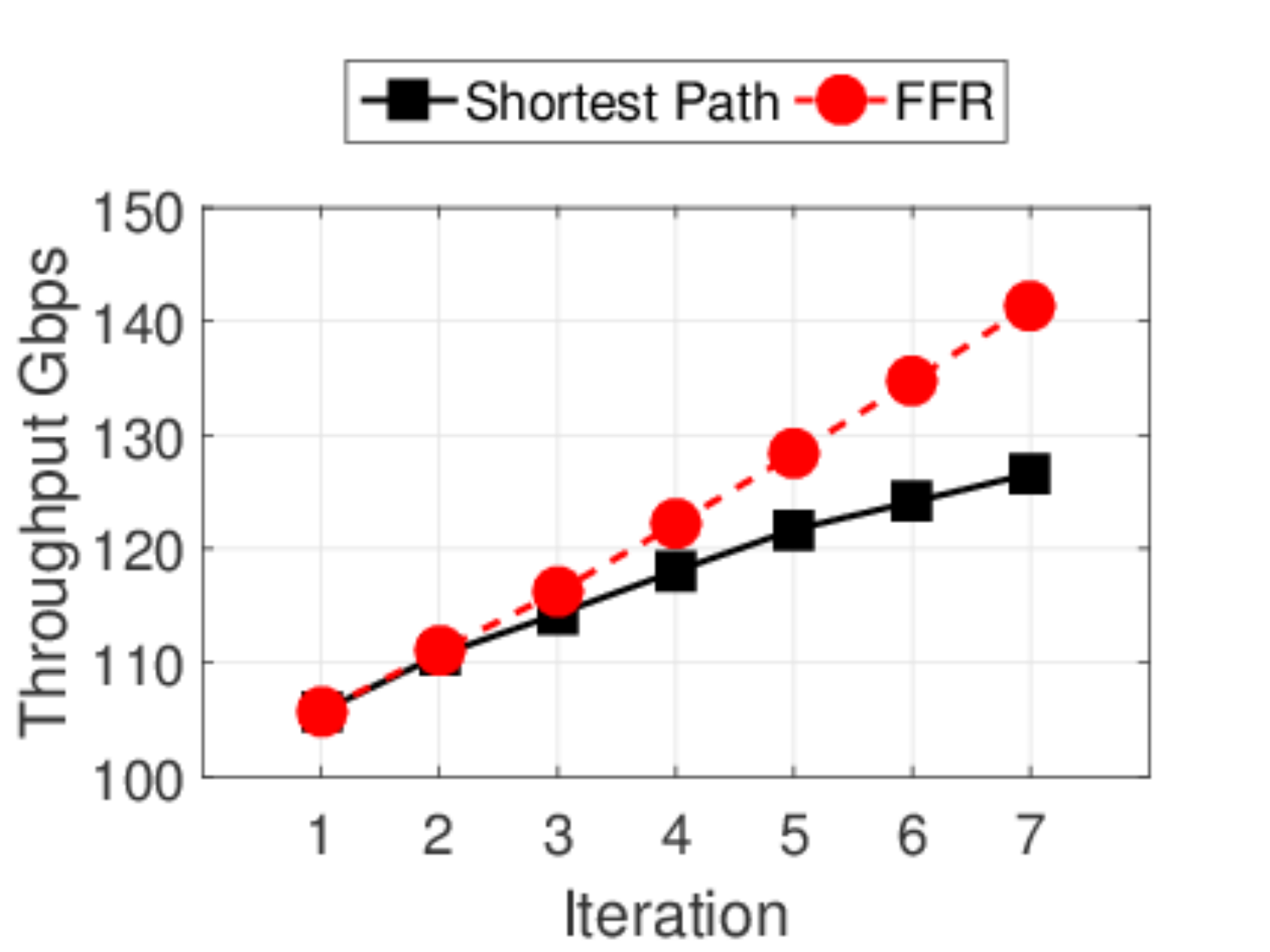}
  \caption{Scenario 2}
  \label{fig:fig86th}
\end{subfigure}
\begin{subfigure}{0.49\columnwidth}
  \centering
  \includegraphics[width=1\linewidth]{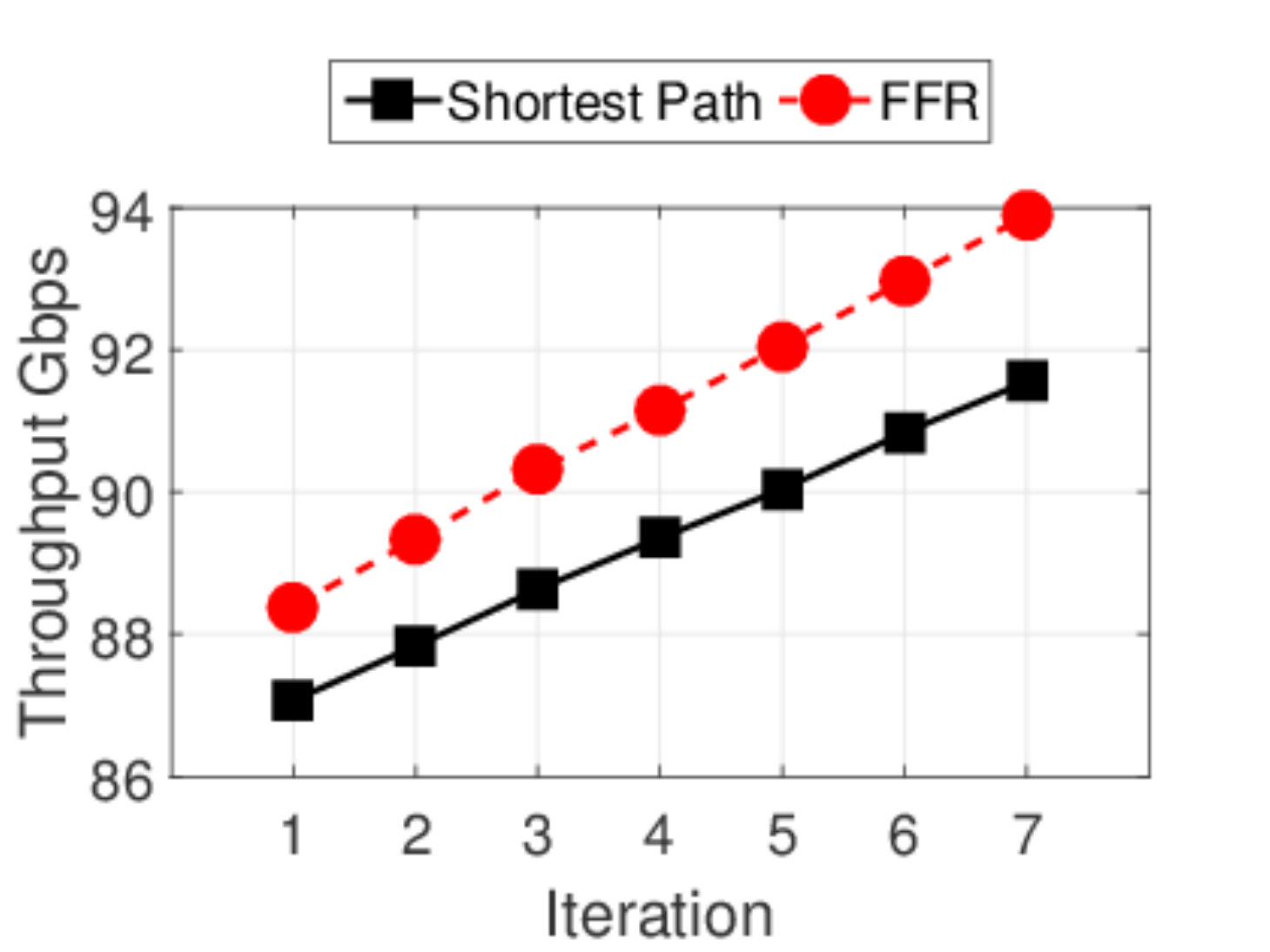}
  \caption{Scenario 3}
  \label{fig:fig71th}
\end{subfigure}
\begin{subfigure}{0.49\columnwidth}
  \centering
  \includegraphics[width=1\linewidth]{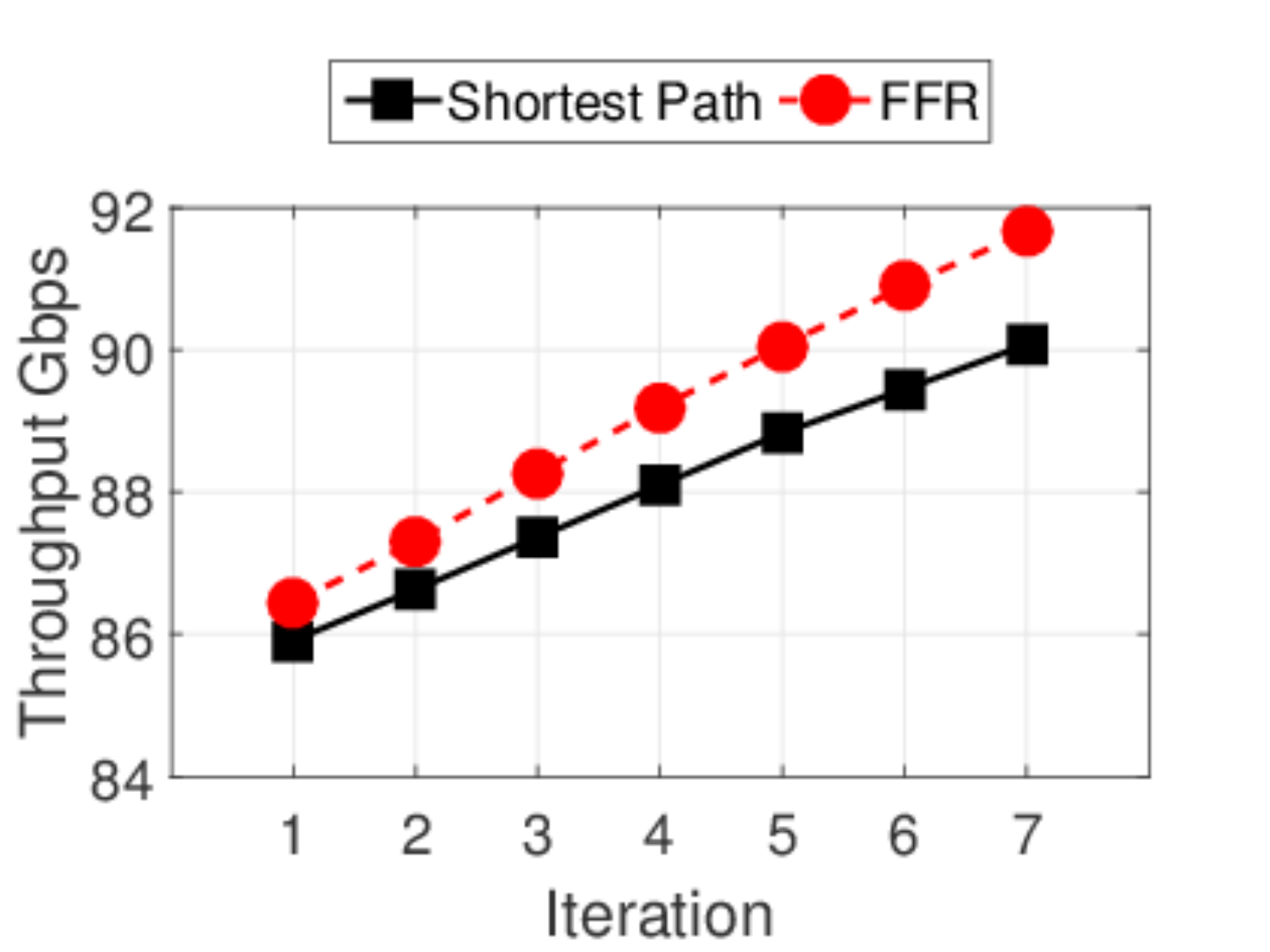}
  \caption{Scenario 4}
  \label{fig:fig70th}
\end{subfigure}%
\caption{Network Throughput.}
\label{fig:Throughput}
\end{figure}

In order to analyze the impact of the proposed scheme on the network throughput, Fig. \ref{fig:Throughput} depicts the network throughput versus time slots. In each time slot the size of flows is increased using uniform distribution by a factor of $2$ percent in scenarios 1 and 2, and a factor of $10$ percent in scenarios 3 and 4. For example, in scenario 3, the size of traffic flows is increased at most $10$ percent in each time slot while the average rate of increment is $5$ percent and the minimum rate of increment is $0$.

The proposed scheme considers the impact of flows on each other, therefore it distributes the flows among divers paths. This behaviour, enhances the network throughput significantly. This happens because, in the traditional approaches like shortest path, each flow is routed separately while our scheme uses resource reallocation to prevent resource partitioning and simultaneously prevents congestion. Based on these results, the impact of the proposed scheme is increased by increasing the traffic demands. One of the main reason is that increasing the demands increases the probability of resource partitioning in traditional approaches. Another reason is that the traditional approaches do not consider the dynamicity of demands while our scheme exploits a light weight reconfiguration to manage the dynamic nature of traffic.

\subsection{Link Utilization Results}
\begin{figure}[!htbp]
    \centering
    \begin{subfigure}{0.49\columnwidth}
      \centering
      \includegraphics[width=1\linewidth]{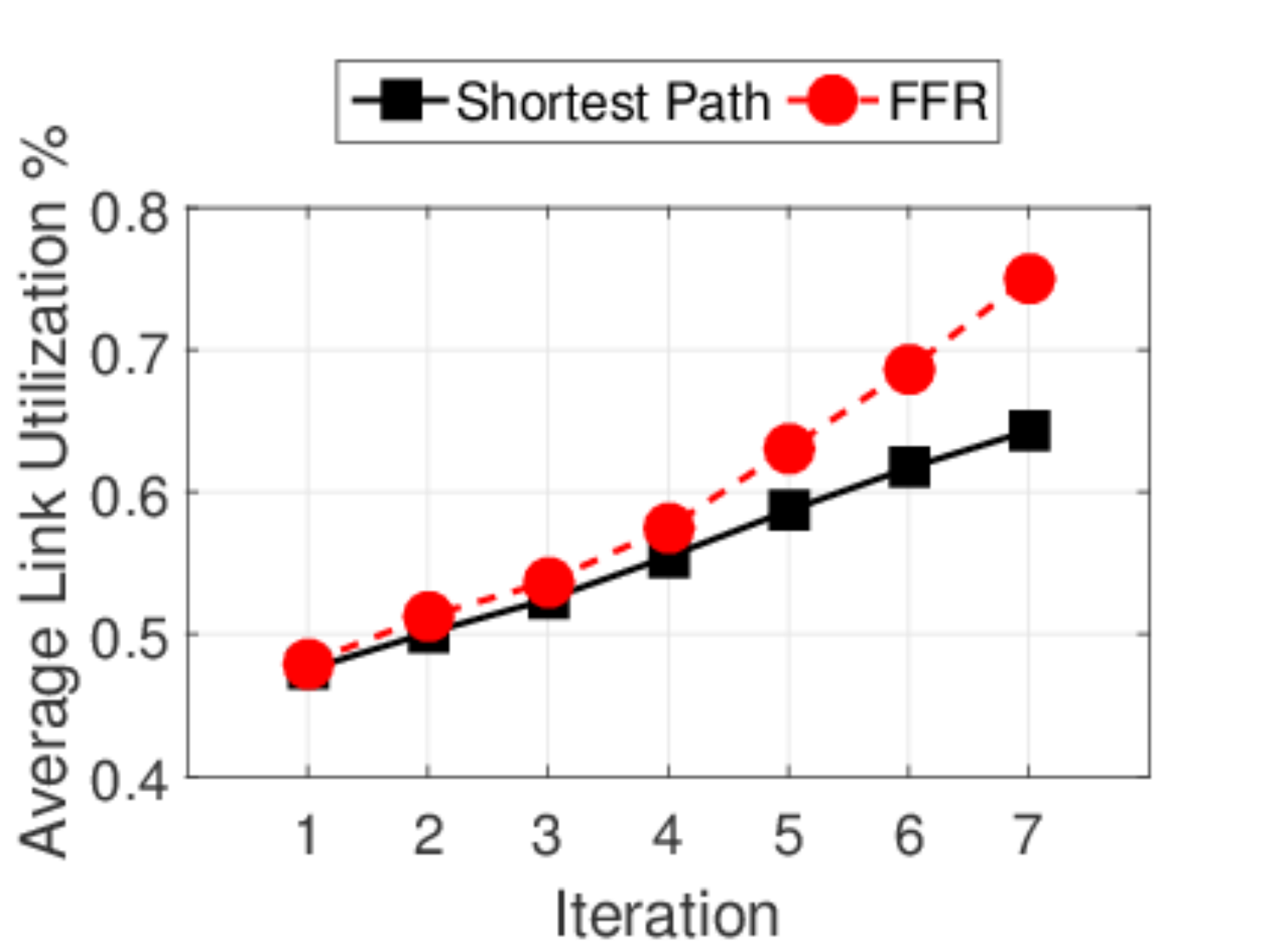}
      \caption{Scenario 1}
      \label{fig:fig84ln}
    \end{subfigure}
    \begin{subfigure}{0.49\columnwidth}
      \centering
      \includegraphics[width=1\linewidth]{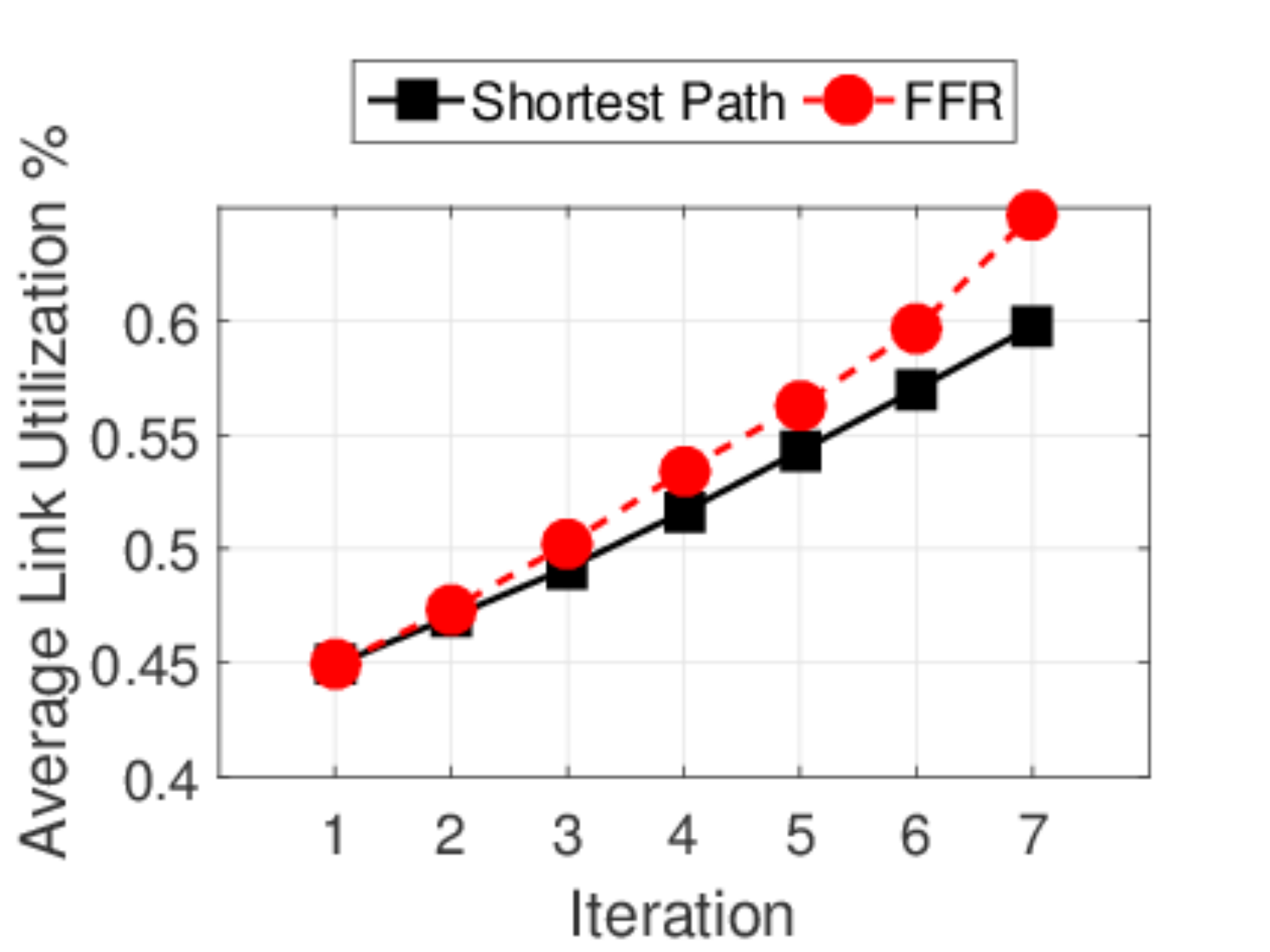}
      \caption{Scenario 2}
      \label{fig:fig86ln}
    \end{subfigure}
    \begin{subfigure}{0.49\columnwidth}
      \centering
      \includegraphics[width=1\linewidth]{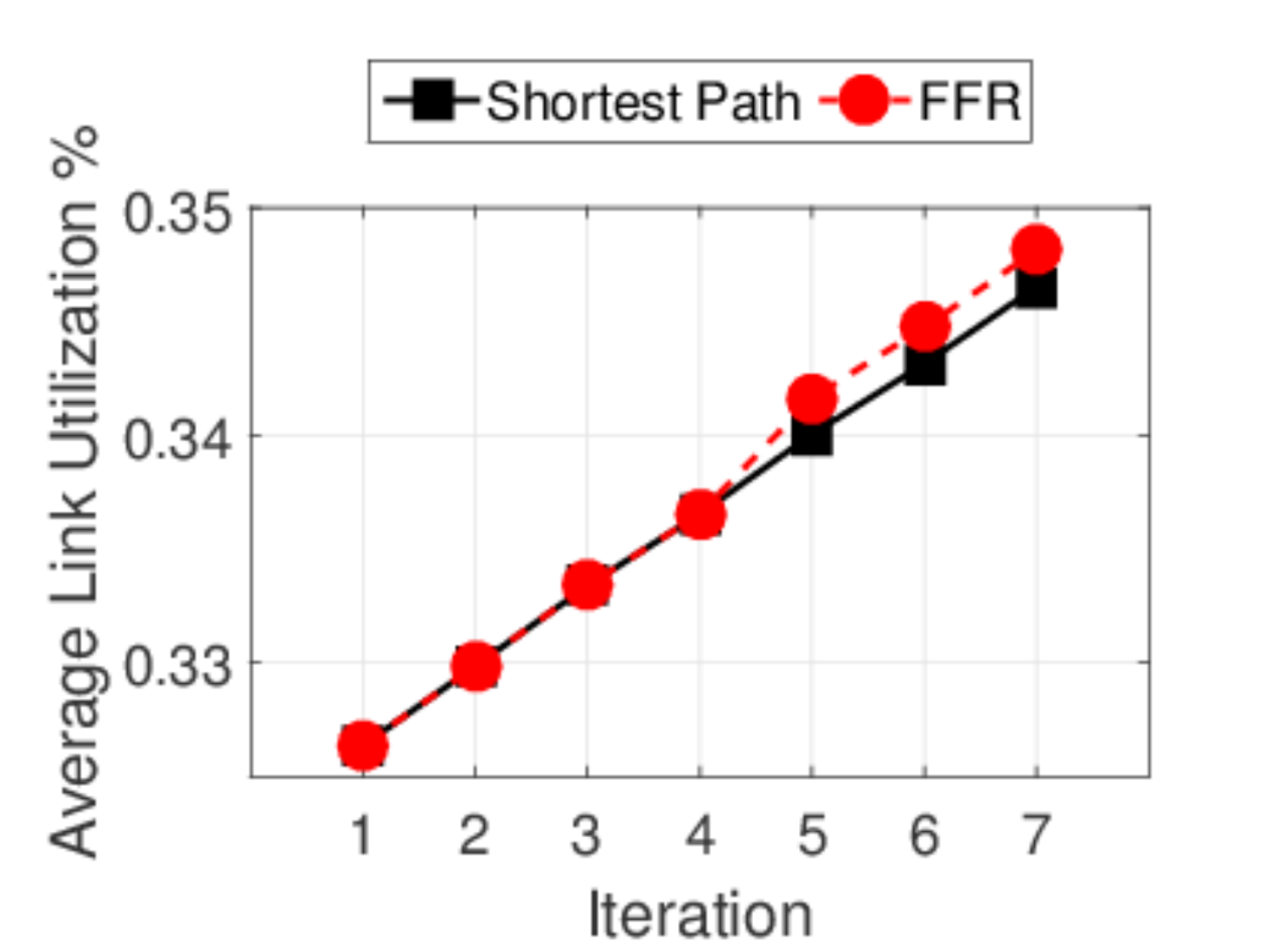}
      \caption{Scenario 3}
      \label{fig:fig71ln}
    \end{subfigure}
    \begin{subfigure}{0.49\columnwidth}
      \centering
      \includegraphics[width=1\linewidth]{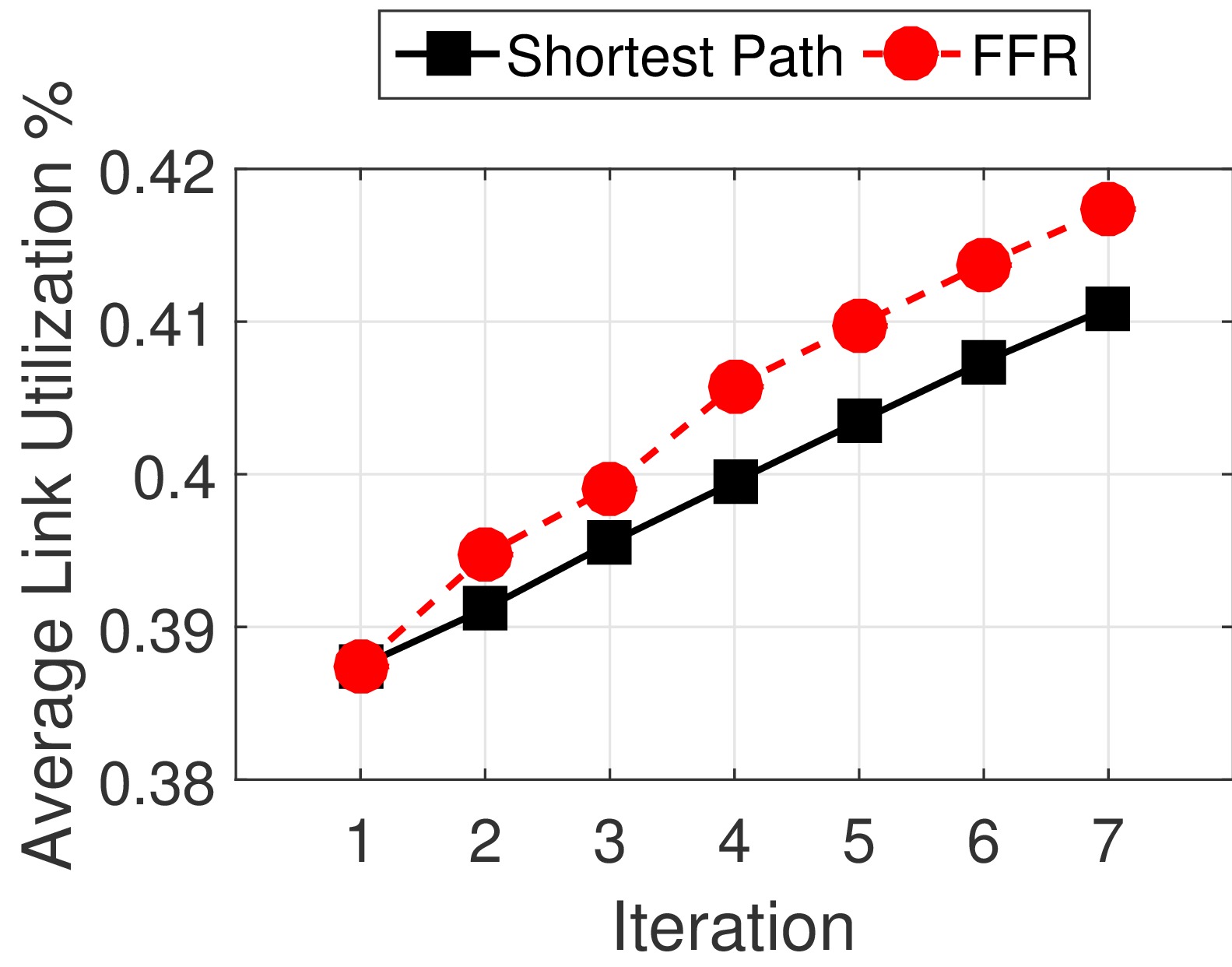}
      \caption{Scenario 4}
      \label{fig:fig70ln}
    \end{subfigure}%
    \caption{Average Link Utilization.}
    \label{fig:LinkUtilization}
    \end{figure}

In order to provide a comprehensive analysis, we investigate the impact of the proposed scheme on the average links utilization in different traffic scenarios. Fig. \ref{fig:LinkUtilization} depicts the average links utilization versus the time slots. As can be seen, the results of both approaches are similar in low traffic demands, however, increasing the traffic demand causes congestion in the shortest path and decreases the average links utilization.
    
More precisely, since there is no congestion while the amount of traffic demands are sufficiently lower than the resources, the result of both approaches is similar. However, increasing the traffic demand in all test cases causes the average link utilization of the proposed scheme to grow higher than the results of shortest path. This happens because the throughput of the proposed scheme is higher than shortest path, consequently the total amount of the traffic loaded on links is higher. 

\subsection{Path Length Results}
\begin{figure}[!htbp]
    \centering
    \begin{subfigure}{0.49\columnwidth}
      \centering
      \includegraphics[width=1\linewidth]{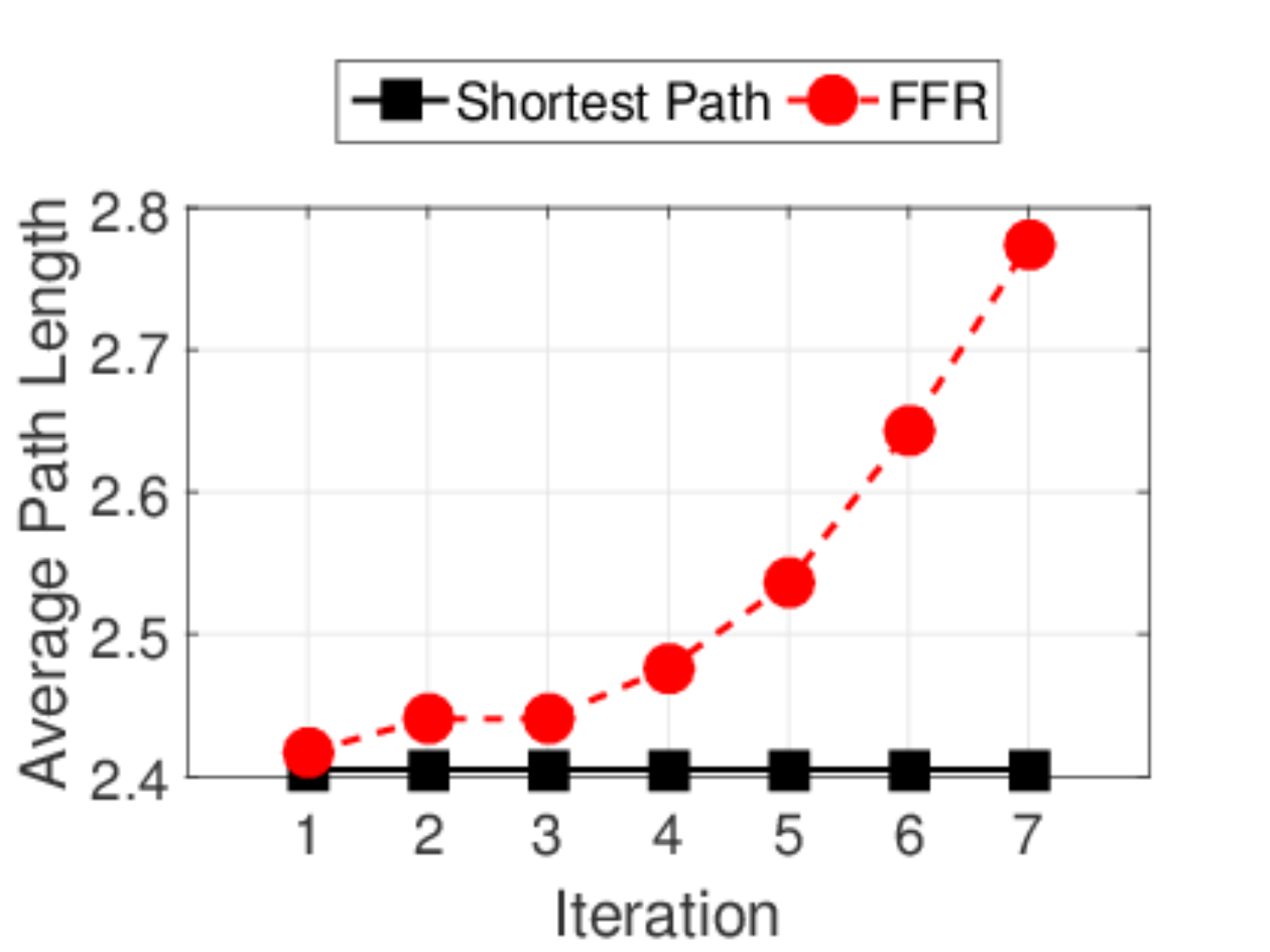}
      \caption{Scenario 1}
      \label{fig:fig84pt}
    \end{subfigure}
    \begin{subfigure}{0.49\columnwidth}
      \centering
      \includegraphics[width=1\linewidth]{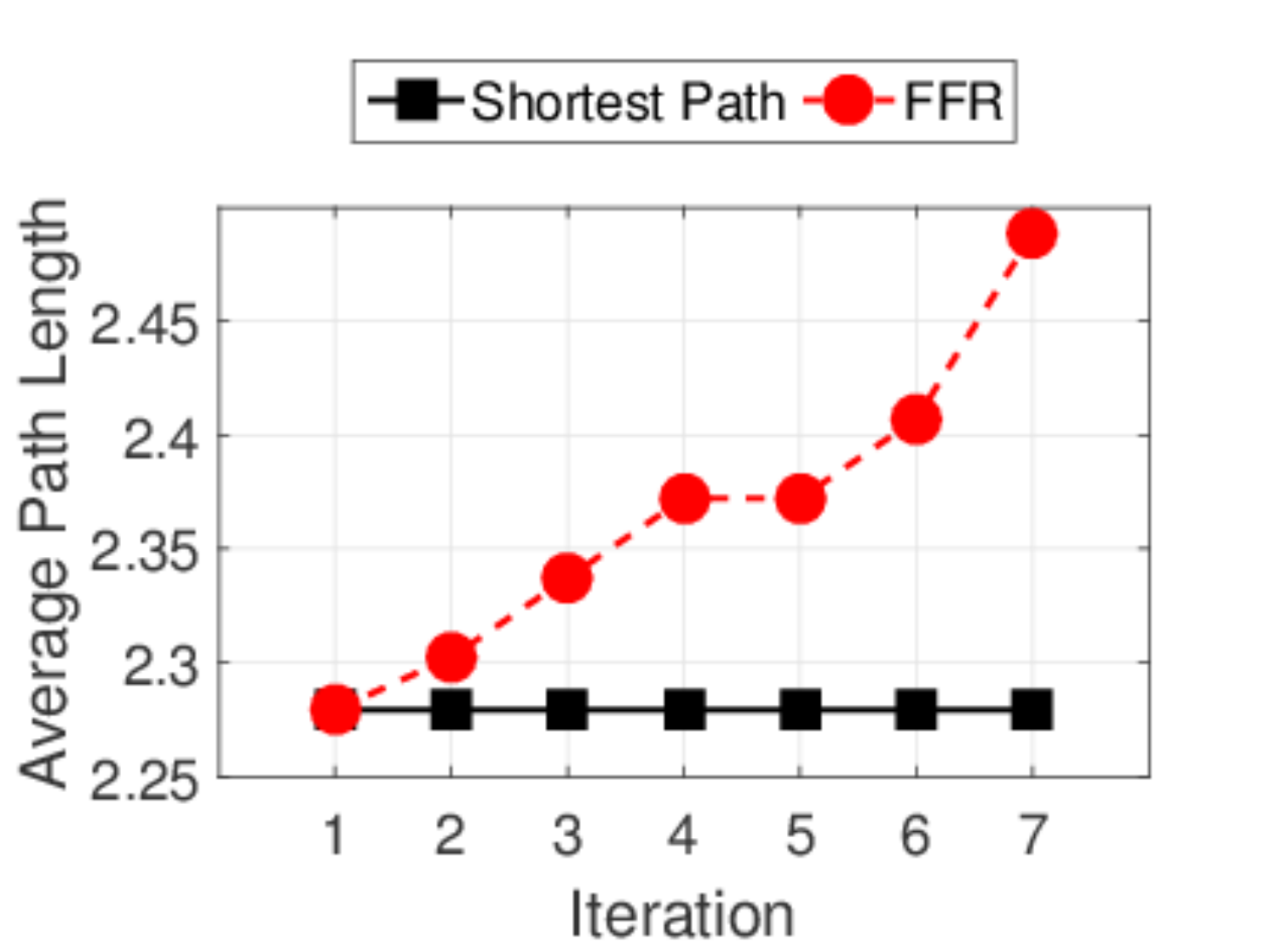}
      \caption{Scenario 2}
      \label{fig:fig86pt}
    \end{subfigure}
    \begin{subfigure}{0.49\columnwidth}
      \centering
      \includegraphics[width=1\linewidth]{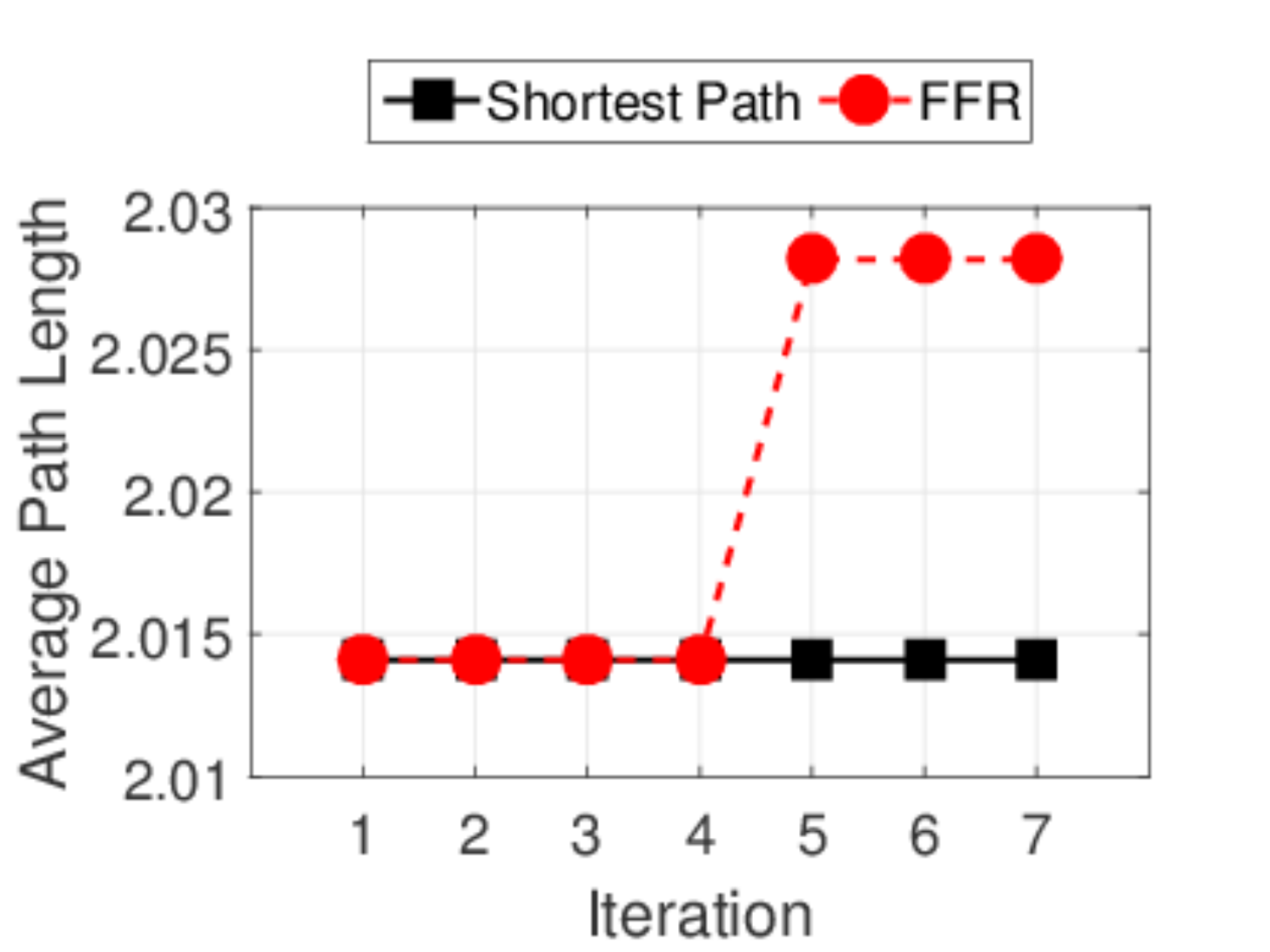}
      \caption{Scenario 3}
      \label{fig:fig71pt}
    \end{subfigure}
    \begin{subfigure}{0.49\columnwidth}
      \centering
      \includegraphics[width=1\linewidth]{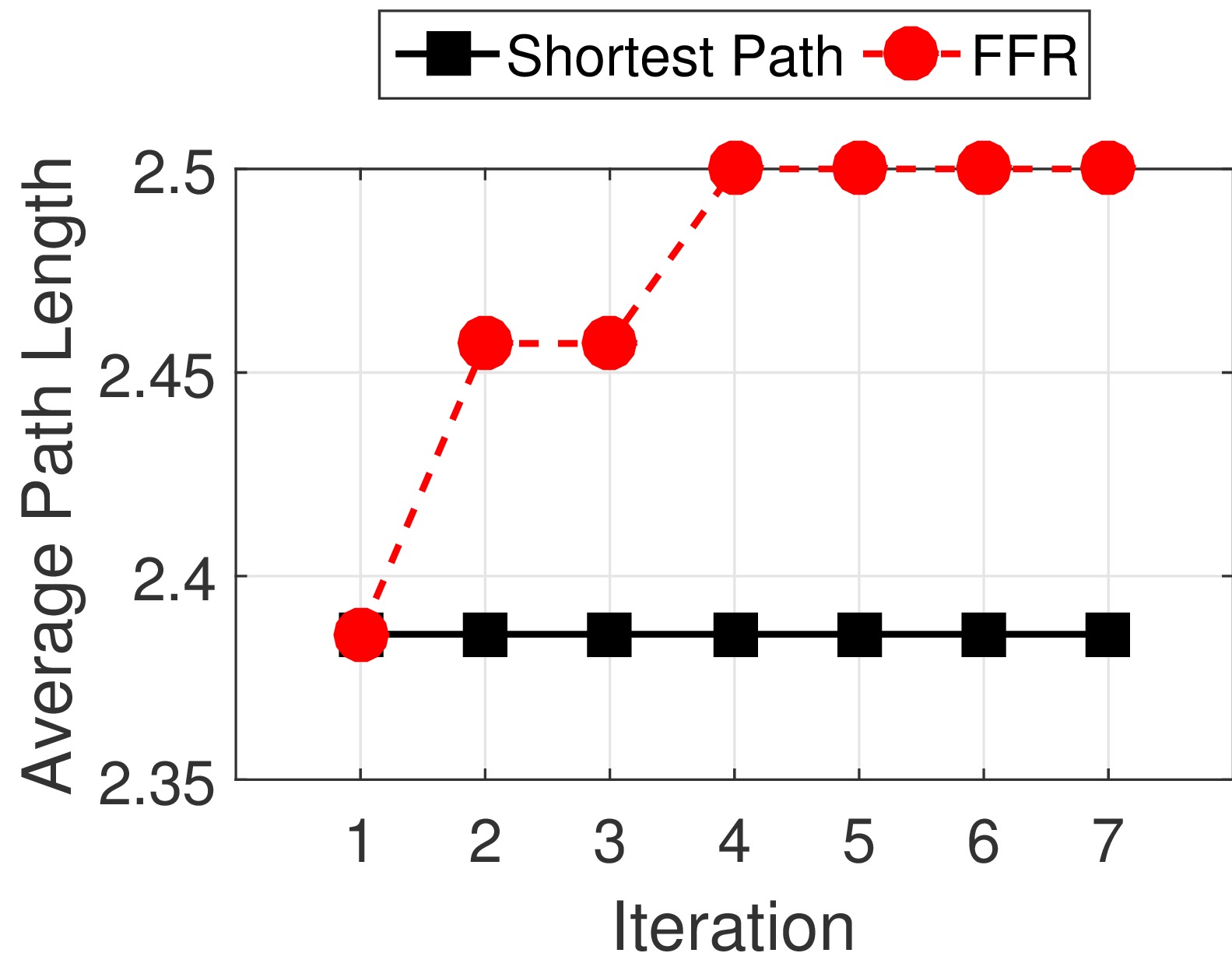}
      \caption{Scenario 4}
      \label{fig:fig70pt}
    \end{subfigure}%
    \caption{Average Path Length.}
    \label{fig:PathLength}
    \end{figure}

Fig. \ref{fig:PathLength} depicts the average path length versus the time slots. Increasing the traffic demands, makes the proposed scheme to use several paths with different length to prevent the network congestion. Additionally, since our scheme considers the impact of flows on each other, it uses paths with minimum common links. Therefore, the average path length increases in compared with the traditional approaches. 
    
Considering Fig. \ref{fig:PathLength} and Fig. \ref{fig:Throughput}, although the network throughput of the proposed scheme is increased significantly in compared with shortest path, the average path length is similar in the both schemes. In other words, although the proposed scheme uses divers paths to reduce the packet loss, the average end-to-end propagation delay is still comparable with the shortest path. It should be mentioned that the formulation checks the end-to-end delay and assigns LSPs to the flows in a way that the path delay is less than the maximum tolerable delay of the flows.

\section{Conclusion}\label{conclusion}
In this paper, a traffic engineering architecture for the hybrid networks of SDN and MPLS introduced. The proposed scheme not only exploits the flexibility of SDN-based approaches but also is applicable on the existing MPLS networks by adding a few number of low-cost OpenFlow-enabled switches. To this end, we mathematically formulated two optimization problems: a) the problem of LSPs re-configuration in MPLS networks when there is a central controller as the PCE element, and b) the problem of flow-level resource re-allocation. The simulation results shows that the proposed scheme increases the network throughput and reduces the total packet loss significantly. Future works will be dedicated on proposing heuristic approaches to consider the energy consumption of the network.

\bibliographystyle{IEEEtran}
\bibliography{IEEEabrv,scholar}

\begin{thebibliography}{10}
\providecommand{\url}[1]{#1}
\csname url@samestyle\endcsname
\providecommand{\newblock}{\relax}
\providecommand{\bibinfo}[2]{#2}
\providecommand{\BIBentrySTDinterwordspacing}{\spaceskip=0pt\relax}
\providecommand{\BIBentryALTinterwordstretchfactor}{4}
\providecommand{\BIBentryALTinterwordspacing}{\spaceskip=\fontdimen2\font plus
\BIBentryALTinterwordstretchfactor\fontdimen3\font minus
  \fontdimen4\font\relax}
\providecommand{\BIBforeignlanguage}[2]{{%
\expandafter\ifx\csname l@#1\endcsname\relax
\typeout{** WARNING: IEEEtran.bst: No hyphenation pattern has been}%
\typeout{** loaded for the language `#1'. Using the pattern for}%
\typeout{** the default language instead.}%
\else
\language=\csname l@#1\endcsname
\fi
#2}}
\providecommand{\BIBdecl}{\relax}
\BIBdecl

\bibitem{tu2014splicing}
X.~Tu, X.~Li, J.~Zhou, and S.~Chen, ``Splicing mpls and openflow tunnels based
  on sdn paradigm,'' in \emph{Cloud Engineering (IC2E), 2014 IEEE International
  Conference on}.\hskip 1em plus 0.5em minus 0.4em\relax IEEE, 2014, pp.
  489--493.

\bibitem{KirkpatricSDN}
K.~Kirkpatrick, ``Software-defined networking,'' \emph{Communications of the
  ACM}, vol.~56, pp. 16--19, 2013.

\bibitem{SDNFirstPaper}
N.~McKeown, ``Software-defined networking,'' \emph{INFOCOM keynote talk},
  vol.~17, pp. 30--32, 2009.

\bibitem{OpenFlowProtocol}
N.~McKeown, T.~Anderson, H.~Balakrishnan, G.~Parulkar, L.~Peterson, J.~Rexford,
  S.~Shenker, and J.~Turner, ``Openflow: enabling innovation in campus
  networks,'' \emph{ACM SIGCOMM Computer Communication Review}, vol.~38, pp.
  69--74, 2008.

\bibitem{FarhadiSDNSurv}
H.~Farhady, H.~Lee, and A.~Nakao, ``Software-defined networking: A survey,''
  \emph{Computer Networks}, vol.~81, pp. 79--95, 2015.

\bibitem{GholCong}
M.~Gholami and B.~Akbari, ``Congestion control in software defined data center
  networks through flow rerouting,'' in \emph{Proceeding of the Electrical
  Engineering (ICEE), 2015 23rd Iranian Conference on}.\hskip 1em plus 0.5em
  minus 0.4em\relax Tehran, Iran: IEEE, 2015, pp. 654--657.

\bibitem{TajikQRTP}
M.~M. Tajiki, B.~Akbari, and N.~Mokari, ``{QRTP}: Qo{S}-aware resource
  reallocation based on traffic prediction in software defined cloud
  networks,'' in \emph{Proceeding of the Telecommunications (IST), 2016 8th
  International Symposium on}.\hskip 1em plus 0.5em minus 0.4em\relax Tehran,
  Iran: IEEE, 2016, pp. 527--532.

\bibitem{AkyildizSDNRoadMap}
I.~F. Akyildiz, A.~Lee, P.~Wang, M.~Luo, and W.~Chou, ``A roadmap for traffic
  engineering in {SDN}-{OpenFlow} networks,'' \emph{Computer Networks},
  vol.~71, pp. 1--30, 2014.

\bibitem{TajikSDTE}
K.~Marzieh, M.~M. Tajiki, and B.~Akbari, ``{SDTE}:software defined traffic
  engineering for improving data center network utilization,''
  \emph{International Journal Information and Communication Technology
  Research}, vol.~8, pp. 15--24, 2016.

\bibitem{tajiki2017MDPI}
M.~M. Tajiki, B.~Akbari, M.~Shojafar, and N.~Mokari, ``Joint qos and congestion
  control based on traffic prediction in sdn,'' \emph{Applied Sciences},
  vol.~7, no.~12, p. 1265, 2017.

\bibitem{salsanoHybridIPSDN}
S.~Salsano, P.~L. Ventre, F.~Lombardo, G.~Siracusano, M.~Gerola, E.~Salvadori,
  M.~Santuari, M.~Campanella, and L.~Prete, ``Hybrid {IP/SDN} networking: open
  implementation and experiment management tools,'' \emph{IEEE Transactions on
  Network and Service Management}, vol.~13, pp. 138--153, 2016.

\bibitem{lopexTwardTransSDN}
V.~Lopez, L.~M. Contreras, O.~G. de~Dios, and J.~P.~F. Palacios, ``Towards a
  transport {SDN} for carriers networks: An evolutionary perspective,'' in
  \emph{Proceeding of the Networks and Optical Communications (NOC), 2016 21st
  European Conference on}.\hskip 1em plus 0.5em minus 0.4em\relax Lisbon,
  Portugal: IEEE, 2016, pp. 52--57.

\bibitem{AguadoABNO}
A.~Aguado, V.~L{\'o}pez, J.~Marhuenda, {\'O}.~G. de~Dios, and J.~P.
  Fern{\'a}ndez-Palacios, ``{ABNO}: A feasible sdn approach for multivendor ip
  and optical networks [invited],'' \emph{Journal of Optical Communications and
  Networking}, vol.~7, pp. A356--A362, 2015.

\bibitem{SgambelSDNPCE}
A.~Sgambelluri, F.~Paolucci, A.~Giorgetti, F.~Cugini, and P.~Castoldi, ``{SDN}
  and {PCE} implementations for segment routing,'' in \emph{Proceeding of the
  Networks and Optical Communications-(NOC), 2015 20th European Conference
  on}.\hskip 1em plus 0.5em minus 0.4em\relax London, UK: IEEE, 2015, pp. 1--4.

\bibitem{DasMPLSOPN}
S.~Das, A.~R. Sharafat, G.~Parulkar, and N.~McKeown, ``{MPLS} with a simple
  {OPEN} control plane,'' in \emph{Proceeding of the Optical Fiber
  Communication Conference and Exposition (OFC/NFOEC), 2011 and the National
  Fiber Optic Engineers Conference}.\hskip 1em plus 0.5em minus 0.4em\relax Los
  Angeles, USA: IEEE, 2011, pp. 1--3.

\bibitem{HuiHybnet}
H.~Lu, N.~Arora, H.~Zhang, C.~Lumezanu, J.~Rhee, and G.~Jiang, ``Hybnet:
  Network manager for a hybrid network infrastructure,'' in \emph{Proceedings
  of the Industrial Track of the 13th ACM/IFIP/USENIX International Middleware
  Conference}.\hskip 1em plus 0.5em minus 0.4em\relax New York, USA: ACM, 2013,
  p.~6.

\bibitem{katovHybridSDN}
A.~N. Katov, A.~Mihovska, and N.~R. Prasad, ``Hybrid {SDN} architecture for
  resource consolidation in {MPLS} networks,'' in \emph{Proceeding of the
  Wireless Telecommunications Symposium (WTS), 2015}.\hskip 1em plus 0.5em
  minus 0.4em\relax California, USA: IEEE, 2015, pp. 1--8.

\bibitem{tajiki2017joint}
M.~M. Tajiki, S.~Salsano, M.~Shojafar, L.~Chiaraviglio, and B.~Akbari, ``Joint
  energy efficient and qos-aware path allocation and vnf placement for service
  function chaining,'' \emph{arXiv preprint arXiv:1710.02611}, 2017.

\bibitem{GuoSDNOSPF}
Y.~Guo, Z.~Wang, X.~Yin, X.~Shi, and J.~Wu, ``Traffic engineering in {SDN/OSPF}
  hybrid network,'' in \emph{Proceeding of the Network Protocols (ICNP), 2014
  IEEE 22nd International Conference on}.\hskip 1em plus 0.5em minus
  0.4em\relax California, USA: IEEE, 2014, pp. 563--568.

\bibitem{tajiki2018energy}
M.~M. Tajiki, S.~Salsano, M.~Shojafar, L.~Chiaraviglio, and B.~Akbari,
  ``Energy-efficient path allocation heuristic for service function chaining,''
  in \emph{Proceedings of the 2018 21th Conference on Innovations in Clouds,
  Internet and Networks (ICIN), Paris, France}, 2018, pp. 20--22.

\bibitem{CariaDivid}
M.~Caria, T.~Das, and A.~Jukan, ``Divide and conquer: Partitioning {OSPF}
  networks with {SDN},'' in \emph{Proceeding of the Integrated Network
  Management (IM), 2015 IFIP/IEEE International Symposium on}.\hskip 1em plus
  0.5em minus 0.4em\relax Ottawa, USA: IEEE, 2015, pp. 467--474.

\bibitem{VissicchioSafeUpdate}
S.~Vissicchio, L.~Vanbever, L.~Cittadini, G.~G. Xie, and O.~Bonaventure, ``Safe
  update of hybrid sdn networks,'' \emph{IEEE/ACM Transactions on Networking},
  2017.

\bibitem{HongIncr}
D.~K. Hong, Y.~Ma, S.~Banerjee, and Z.~M. Mao, ``Incremental deployment of
  {SDN} in hybrid enterprise and {ISP} networks,'' in \emph{Proceedings of the
  Symposium on SDN Research}.\hskip 1em plus 0.5em minus 0.4em\relax Santa
  Clara, USA: ACM, 2016, p.~1.

\bibitem{berde2014onos}
P.~Berde, M.~Gerola, J.~Hart, Y.~Higuchi, M.~Kobayashi, T.~Koide, B.~Lantz,
  B.~O'Connor, P.~Radoslavov, W.~Snow \emph{et~al.}, ``Onos: towards an open,
  distributed sdn os,'' in \emph{Proceedings of the third workshop on Hot
  topics in software defined networking}.\hskip 1em plus 0.5em minus
  0.4em\relax ACM, 2014, pp. 1--6.

\bibitem{RFC5440}
J.~Vasseur and J.~Le~Roux, ``{RFC} 5440--path computation element ({PCE})
  communication protocol ({PCEP}),'' \emph{Internet Engineering Task Force
  (IETF)}, 2009.

\bibitem{OnosController}
P.~Berde, M.~Gerola, J.~Hart, Y.~Higuchi, M.~Kobayashi, T.~Koide, B.~Lantz,
  B.~O'Connor, P.~Radoslavov, W.~Snow \emph{et~al.}, ``{ONOS}: towards an open,
  distributed {SDN OS},'' in \emph{Proceedings of the third workshop on Hot
  topics in software defined networking}.\hskip 1em plus 0.5em minus
  0.4em\relax Chicago, USA: ACM, 2014, pp. 1--6.

\bibitem{OpenDaylightController}
J.~Medved, R.~Varga, A.~Tkacik, and K.~Gray, ``Opendaylight: Towards a
  model-driven {SDN} controller architecture,'' in \emph{Proceeding of the A
  World of Wireless, Mobile and Multimedia Networks (WoWMoM), 2014 IEEE 15th
  International Symposium on}.\hskip 1em plus 0.5em minus 0.4em\relax Sydney,
  Australia: IEEE, 2014, pp. 1--6.

\bibitem{QNR}
M.~M. Tajiki, B.~Akbari, and N.~Mokari, ``Optimal {QoS}-aware network
  reconfiguration in software defined cloud data centers,'' \emph{Computer
  Networks}, vol. 120, pp. 71--86, 2017.

\bibitem{tajiki2018CECT}
M.~Tajiki, B.~Akbari, M.~Shojafar, S.~H. Ghasemi, M.~Barazandeh, N.~Mokari,
  L.~Chiaraviglio, and M.~Zink, ``{CECT:} computationally efficient
  congestion-avoidance and traffic engineering in software-defined cloud data
  centers,'' \emph{arXiv preprint arXiv:1710.03611}, 2018.

\bibitem{CrtngAndMngDyn}
H.~Hasan, J.~Cosmas, Z.~Zaharis, P.~Lazaridis, and S.~Khwandah, ``Creating and
  managing dynamic {MPLS} tunnel by using {SDN} notion,'' in \emph{Proceeding
  of the Telecommunications and Multimedia (TEMU), 2016 International
  Conference on}.\hskip 1em plus 0.5em minus 0.4em\relax Crete, Greece: IEEE,
  2016, pp. 1--8.

\end{thebibliography}

\newpage

\end{document}